\documentclass[a4paper, twocolumn, floatfix, 11pt, accepted=2023-04-11]{quantumarticle}
\pdfoutput=1
\usepackage[utf8]{inputenc}
\usepackage[english]{babel}
\usepackage[T1]{fontenc}
\usepackage{amsmath}
\usepackage{hyperref}
\usepackage{xr}

\usepackage{tikz}
\usepackage{lipsum}
\usepackage{amsmath,amssymb,amsfonts}
\usepackage{graphicx}
\usepackage{textcomp}
\usepackage{xcolor}

\usepackage{comment}
\usepackage{physics}
\usepackage{bm}
\usepackage{hyperref} 
\usepackage[normalem]{ulem}
\usepackage{csquotes}
\usepackage{float}

\usepackage{tabularx,booktabs}
\usepackage[margin=10pt,font=small, labelsep=endash]{caption}
\newcolumntype{C}{>{\centering\arraybackslash}X} 

\usepackage{tikz}
  \usetikzlibrary{positioning}
  \usetikzlibrary{quantikz}
\newcommand{\border}[2]{\gategroup[#1,steps=#2,style={dashed,rounded corners,fill=blue!0, inner xsep=0pt, inner ysep = 0pt}, background]{{$\mathds{1}$}}}

\usepackage{dsfont}

\usepackage{physics}

\newcommand{\circuit}[1]{\textsc{C}_{\textsc{#1}}}

\newcommand{\fmap}{\mathcal{F}}
\newcommand{\varans}{V}

\hypersetup{
    colorlinks,
    linkcolor={red!50!black},
    citecolor={blue!50!black},
    urlcolor={blue!80!black}
}

\usepackage{algorithm}
\usepackage{algpseudocode}

\usepackage[numbers]{natbib}

\definecolor{forestgreen}{rgb}{0.13, 0.55, 0.13}
\newcommand{\old}[1]{}
\newcommand{\new}[1]{#1}

\begin{document}

\title{Entanglement entropy production in Quantum Neural Networks}

\author{Marco Ballarin}
\affiliation{These authors contributed equally to this work}
\affiliation{Dipartimento di Fisica e Astronomia "G. Galilei", via Marzolo 8, I-35131, Padova, Italy}
\affiliation{INFN, Sezione di Padova, via Marzolo 8, I-35131, Padova, Italy}
\email{marco.ballarin.6@phd.unipd.it}

\author{Stefano Mangini}
\affiliation{These authors contributed equally to this work}
\affiliation{Dipartimento di Fisica, Università di Pavia, Via Bassi 6, I-27100, Pavia, Italy}
\affiliation{INFN Sezione di Pavia, Via Bassi 6, I-27100, Pavia, Italy}

\author{Simone Montangero}
\affiliation{Dipartimento di Fisica e Astronomia "G. Galilei", via Marzolo 8, I-35131, Padova, Italy}
\affiliation{INFN, Sezione di Padova, via Marzolo 8, I-35131, Padova, Italy}
\affiliation{Padua Quantum Technologies Research Center, Università degli Studi di Padova}

\author{Chiara Macchiavello}
\affiliation{Dipartimento di Fisica, Università di Pavia, Via Bassi 6, I-27100, Pavia, Italy}
\affiliation{INFN Sezione di Pavia, Via Bassi 6, I-27100, Pavia, Italy}
\affiliation{CNR-INO - Largo E. Fermi 6, I-50125, Firenze, Italy}

\author{Riccardo Mengoni}
\affiliation{CINECA Quantum Computing Lab,Via Magnanelli, 6/3, 40033 Casalecchio di Reno, Bologna, Italy}

\maketitle
\begin{abstract}
Quantum Neural Networks (QNN) are considered a candidate for achieving quantum advantage in the Noisy Intermediate Scale Quantum computer (NISQ) era. Several QNN architectures have been proposed and successfully tested on benchmark datasets for machine learning.
However, quantitative studies of the QNN-generated entanglement have been investigated  only for up to few qubits. Tensor network methods allow to emulate quantum circuits with a large number of qubits in a wide variety of scenarios.
Here, we employ matrix product states to characterize recently studied QNN architectures with random parameters up to fifty qubits showing that their entanglement, measured in terms of entanglement entropy between qubits, tends to that of Haar distributed random states as the depth of the QNN is increased. We certify the randomness of the quantum states also by measuring the expressibility of the circuits, as well as using tools from random matrix theory. We show a universal behavior for the rate at which entanglement is created in any given QNN architecture, and consequently introduce a new measure to characterize the entanglement production in QNNs: the entangling speed. Our results characterise the entanglement properties of quantum neural networks, and provides new evidence of the rate at which these approximate random unitaries.
\end{abstract}

\section{Introduction}
Nowadays quantum computing is a well-established research field where quantum phenomena like superposition and entanglement are exploited in order to process information, possibly more efficiently than standard classical data processing~\cite{NielsenChuang}. The aim of quantum computing is to devise quantum algorithms capable of generating a target quantum state representing the solution of a given problem. 
In the last decade, the community has put a large effort into the realization of hardware able to perform quantum computation.

Accompanying the rise of quantum computing, another research area, namely Machine Learning (ML), has gained a lot of popularity. We live undoubtedly in  the  era of big data, where information is collected by the most disparate devices. In this context, ML constitutes a set of techniques for efficiently identifying patterns in huge datasets and for inferring input-output relations in data, even in the case of previously unseen inputs~\cite{Goodfellow2016DeepL, LeCun2015DeepL}. ML proves to be a powerful tool with a wide range of applications: from image classifications~\cite{AlexNet}, over devising playing strategies for complex games~\cite{MasteringGo}, to  controlling nuclear fusion reactors~\cite{DegraveMagneticControlTokamak2022}.

Inspired by some of these outstanding results, a new interdisciplinary research topic that goes by the name of Quantum Machine Learning (QML) has recently begun to combine quantum computing and machine learning techniques in various ways, with the hope of achieving improvements in both fields~\cite{Biamonte2017QML, DunjkoWittekReview2020, CerezoPQC2021Review, ManginiQNN, Bharti2021NIQSReview}. As small-scale quantum devices start to be available~\cite{Preskill2018NISQ}, a new class of quantum procedures called \textit{variational quantum algorithms} have been developed to take advantage of current and near-term quantum hardware, by trading theoretical guarantees of success with feasibility of execution~\cite{CerezoPQC2021Review, Bharti2021NIQSReview, PeruzzoVQE2014}. Generally speaking, a variational quantum  circuit is a hybrid quantum-classical algorithm employing a quantum circuit $U(\bm{\theta})$ that depends on a set of parameters $\bm{\theta}$, which are adjusted in order to minimize a given objective function. While the objective function is evaluated by measuring outcomes of the variational circuit, optimization is performed by a classical iterative optimization algorithm that proposes better candidates for the parameters $\bm{\theta}$, starting from random (or pre-trained) initial values.

Within the domain of variational quantum circuits, quantum versions of neural networks, often referred to as quantum neural networks (QNNs), represent a promising quantum alternative for classical supervised learning~\cite{ManginiQNN, AbbasPowerQNN2021}. An efficient encoding of input data is key to perform computations in a high dimensional (possibly even infinite) Hilbert space. In fact, it is possible to encode classical inputs $\bm{x}$ into a quantum state $\ket{\fmap(\bm{x})}$ using a parameterized quantum circuit (PQC), a procedure which goes by the name of \textit{feature encoding}. Thus, the goal of the feature map $\fmap$ is to map classical vectors to the qubits' Hilbert space. This feature map is accompanied by a layered structure of additional variational PQCs, which are trained in order to solve the desired learning task. Recently, QNNs gained a lot of attention after it was shown that they could be more expressive and efficiently trained than their classical counterparts~\cite{AbbasPowerQNN2021}. Still, the dispute on how to achieve quantum advantages over machine learning is still far from being settled~\cite{Preskill2021Informationtheoretic, Huang2020Power, SchreiberSurrogates2022}. As for classical neural networks, the type of implementation of parameterized quantum circuits has a profound impact on the  QNN performances, both in terms of trainability and classification accuracy~\cite{hubregtsenEvaluationParameterizedQuantum2021, CerezoBarrenLocalCost2021, CongQCNN2019, MeyerSymmetries, SkolikEquivariant}. Thus, characterizing parameterized quantum circuits in terms of their expressibility and entangling capability is key to selecting a good \textit{ansatz}, i.e. structure, for a QNN. 

\begin{figure*}[ht!]
    \centering
    \includegraphics[width = \textwidth]{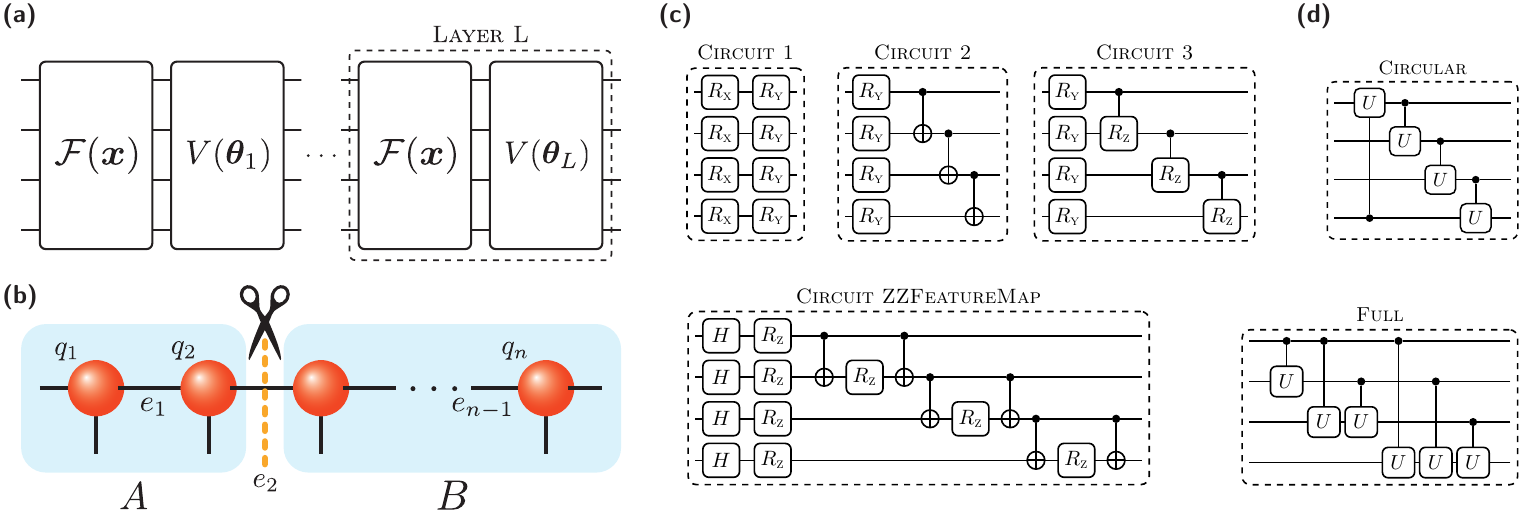}
    \caption{Graphical representation of QNN and MPS. \textbf{(a)} QNN structure with alternating feature map $\fmap$ and variational ansatz $\varans$. Note that the ansatz parameters are different in each layer, while the feature map parameters are the same throughout the whole circuit. \textbf{(b)} MPS diagram. Each sphere is a tensor, representing a qubit $q_j$. The entanglement entropy between bi-partitions $A$ and $B$ is computed by "cutting" the connecting edge $e_j$. \textbf{(c)} Circuits analyzed in the manuscript, depicted with a linear entanglement topology, i.e. entangling gates are only applied between nearest neighbors on a line. \textbf{(d)} Different entanglement topologies: circular, with the first and last qubit of the line connected, and full, where the entangling gates are applied between each pair of qubits (see Appendix \ref{app:entangling_maps} for a clear definition and discussion). When using parameterized two qubits gates, like the controlled rotations in circuit 2, the entanglement maps are generalized to their parameterized version by substituting $X$ gates on the controlled qubits with the corresponding parameterized operation. Note that the circuit templates 2 and \textsc{ZZFeatureMap} are those used in the QNN of~\cite{AbbasPowerQNN2021}, and also that circuits 1, 2 and 3 share similarities with circuits 1, 15, and 13 of~\cite{SimPQCs2019}, respectively.}
    \label{fig:main_figure}
\end{figure*}

Following and expanding the investigation pioneered  in~\cite{SimPQCs2019}, in this work we study the entanglement properties of quantum neural networks initialized with random parameters. We employ methods from the tensor network literature, namely Matrix Product States (MPS), to study the entanglement generated in various QNNs architectures composed of up to 50 qubits. Since MPS are a very powerful tool for simulating quantum systems with bounded entanglement, if a quantum neural network can only access low entangled states, it can be easily simulated, which spoils any hope of achieving a concrete quantum advantage. Thus, using  entanglement entropy among qubits as a figure of merit, we evaluate the entanglement capabilities of some of the most common and promising QNN architectures \cite{AbbasPowerQNN2021, SimPQCs2019}. We consider several QNNs with different combinations of feature maps $\fmap$ and variational forms $\varans$ and perform an extended numerical analysis varying: (\textit{i}) the number of qubits $n$, (\textit{ii}) the number of layers $L$ in the network, (\textit{iii}) the entangling topology of the circuit, (\textit{iv})  the data re-uploading~\cite{Perez2020Reuploading, Schuld2020Encoding} structure being either alternated or sequential. \new{In this respect, we focus our analysis on data re-uploading quantum circuits because, as extensively discussed later, recent results in the quantum machine learning literature highlight the need for such a circuit structure to increase the expressibility of the parametric models implemented by quantum neural networks. Thus, we consider this class of parametrized quantum circuits due to their practical relevance in quantum machine learning tasks. Nonetheless, in Sec.~\ref{sec:MPSSimulation}, we also analyze instances of random quantum circuits where parameters are not shared between layers (hence no data reuploading is used) and show that, as long as entanglement is involved, the results presented in this paper depend primarily on the architecture of the parametric quantum circuit, and not on the presence of shared parameters}. A summary of the circuit templates analyzed in this work is shown in Fig.~\ref{fig:main_figure}.

For all the considered QNNs with nearest neighbour connectivity, as the number of layers $L$ is increased, the entanglement generated inside the circuit grows, eventually reaching a plateau when $L\approx n$, where $n$ is the number of qubits. This behavior is associated with the typical entanglement of a random Haar-distributed quantum state. The choice of the entangling topology (nearest neighbors, circular, or all to all) clearly affects the rate of creation of entanglement in the circuit. We also point out that a careless definition of a full, i.e. all to all, connectivity map can effectively result in a linear nearest-neighbors interaction if unparameterized two qubits gates (\textsc{Cnot}s) are used, something apparently overlooked in the recent literature using this type of ansatz~\cite{AbbasPowerQNN2021, TacchinoVariationalQNN2021, Jaderberg2022SelfSupervised}.
By bounding the entanglement generated by the circuit, we are able to simulate QNNs with MPS up to $n=50$ qubits. It should be stressed that such simulations are exact up to a given number of layers, after which a truncation of the entanglement via MPS is applied. By appropriately normalizing the entanglement produced we show that all the points for a given QNN architecture follow the same curve, independently from the number of qubits. Thus, we exploit this behavior to define a universal figure of merit given the QNN architecture, the entangling speed. This figure of merit characterizes how fast the entanglement is produced by the QNN, with respect to the number of layers $L$.

In addition, we evaluate the expressibility measure of the considered QNNs as defined in~\cite{SimPQCs2019} and argue that the optimality of the QNN introduced in~\cite{AbbasPowerQNN2021} may be related to its good trade-off between mild entanglement production and high expressibility. Finally, we employ tools from random matrix theory, specifically convergence to the Mar\v{c}enko-Pastur distribution, to further characterise the resemblance of the deep enough quantum neural networks to random unitary matrices. At last, we note that differently from~\cite{SimPQCs2019} which bases their analysis on the Meyer-Wallach entanglement measure~\cite{meyerGlobalEntanglementMultiparticle2002}, in this work we make use of the entanglement entropy among subsystems, which allows for a more careful analysis of the entanglement distribution in the system, and it is also readily accessed in an MPS simulation with no computational overhead.

The manuscript is organized as follows. In Sec.~\ref{sec:methods} we review the basis of tensor networks and MPS, and introduce the Von Neumann entropy as an entanglement measure. We then discuss the entanglement entropy properties of random quantum states. We proceed by discussing the most recent results on parameterized quantum circuits and QNNs, especially, on the relation between randomness, trainability, and entanglement found in these circuits.
In Sec.~\ref{sec:Results} we show the results of our analysis for various QNN architectures, and discuss the results in Sec.~\ref{sec:Discussion}. 
Finally, we discuss the implications of our work and possible routes for future investigations in Sec.~\ref{sec:Conclusion}. 

\section{Methods}\label{sec:methods}

\subsection{Tensor Networks and Matrix Product States}\label{sec:MPS}
An $n$-qubit quantum state is defined in a Hilbert space $\mathcal{H}$ of dimension $\mbox{dim}(\mathcal{H})=2^n$. The exponential scaling of $\mathcal{H}$ with $n$ makes the classical description of quantum states an exponentially expensive task. This problem is widely known in many-body quantum physics, and many different techniques have been developed to alleviate the issue, like the Density Matrix Renormalization Group (DMRG) or Tensor Network (TN) techniques~\cite{tensor_anthology, Montangero_book}. 

In this work, we use Tensor Network methods to efficiently describe the $n$-qubit state. In particular, we employ Matrix Product States (MPS), which are a specific tensor network ansatz particularly suited to represent 1-dimensional (i.e. like atoms on a chain, as in Fig.~\ref{fig:main_figure}) quantum states~\cite{EisertEntanglementMPS}. The power of tensor networks lies in the assumption that we are only interested in a \textit{tiny subspace} of the entire Hilbert space, namely the states that display a limited amount of entanglement.

An $n$-qubit pure state $\ket{\psi} \in \mathcal{H}$ can be written as a MPS as follows~\cite{EisertEntanglementMPS}
\begin{equation}
    \label{eq:mps_representation}
    \begin{aligned}
    \ket{\psi} =& \sum_{{s_1,\dots,s_n=0}}^1\, \sum_{{\alpha_1, \dots, \alpha_n = 1}}^\chi  \textbf{M}_{1 \alpha_1}^{[1],s_1}\, \textbf{M}_{\alpha_1  \alpha_2}^{[2],s_2} \cdots \\ 
    &\quad\,\, \textbf{M}_{\alpha_{n-2} \alpha_{n-1}}^{[n-1],s_{n-1}}\, \textbf{M}_{\alpha_{n-1} 1}^{[n], s_n}\, \ket{s_1 s_2 \hdots s_n}\,.
    \end{aligned} 
\end{equation}
Each tensor $\textbf{M}_{\alpha_{i} \alpha_{i+1}}^{[i], s_i}$ is a local description for the $[i]$-th site, which allows one to apply a \textit{local} operator to a certain site without the need to change all the other coefficients. For a fixed $s_i$, $\textbf{M}_{\alpha_i \alpha_{i+1}}^{[i],s_i}$ is a $\chi \times \chi$ complex matrix, meaning that  Eq.~\eqref{eq:mps_representation} is the sum of basis elements weighted by matrix products. The integer $\chi$ is called the MPS \textit{bond dimension}, and a sufficiently high $\chi$ is needed to express a general $\ket{\psi}$ in such form. However, MPS with a \textit{lower} $\chi$ can still encode all the meaningful states, albeit clearly not \textit{all} possible states. In particular, to correctly describe any quantum state the bond dimension needed is $\chi=d^{\lfloor\frac{n}{2}\rfloor}$, where $d$ is the local dimension of the degrees of freedom ($d=2$ for qubits).
We can also efficiently evolve the state under the application of $2$-qubit gates, using an approach known in the literature as time-evolving block decimation~\cite{time_evolution_mps}, and perform measurements. Simulations using MPS are not bounded by the number of qubits in the system, but by the amount of entanglement generated inside it, as we explain in detail in Section \ref{sec:Entanglement}.

Nonetheless, while the use of an MPS simulation imposes some constraints on the maximum entanglement that it is possible to represent, this issue is relevant only for very deep circuits involving many qubits. Indeed, we reliably simulate circuit instances involving up to $n=50$ qubits and moderate depth, which is already sufficient to provide clear insights on the entanglement entropy generated in such circuits. Moreover, as explained below in Sec.~\ref{sec:Entanglement}, during an MPS simulation one has constant access to the singular values of the quantum state, so the entanglement of the state can be calculated on the fly without any computational overhead. Thus, MPS are an effective tool to study the entanglement properties of quantum circuits, especially in regimes that cannot be easily accessed with a full-scale simulation of the statevector of the system.

\subsection{Entanglement measure in Matrix Product States}\label{sec:Entanglement}
Entanglement in quantum states can be evaluated using the so-called Von Neumann \textit{entanglement entropy}. Let $\rho = \dyad{\psi}$ be the quantum state of a system of $n$ qubits, and consider a bipartition $A\,,B$ of such system of qubits $n_A$ and $n_B=n-n_A$ respectively, like the one shown in Fig.~\ref{fig:main_figure}(b). The entanglement entropy of the subsystem $A$ having reduced density matrix $\rho_A = \Tr_B[\rho]$, is defined as
\begin{equation}
\label{eq:ent_entropy}
    S(\rho_A) = -\Tr[\rho_A\log\rho_A]\,,
\end{equation}
and quantifies the amount of entanglement shared between the parties $A$ and its complement $B$. Note that throughout the whole manuscript we consider logarithms in natural base $e$. If $A$ and $B$ are in a product state then $S(\rho_A)=0$, while if the two subsystems share maximal entanglement one has $S(\rho_A) = n_A\,\log(2)$~\cite{EisertEntanglementMPS}. An important property of Eq.~(\ref{eq:ent_entropy}) is that the entanglement entropy of the two subsystems is equal, namely $S(\rho_A) = S(\rho_B)$, as it can be easily checked using the Schmidt decomposition of the pure global state $\rho = \dyad{\psi}$ (see below). 

It turns out that matrix product states are a natural tool to characterize the entanglement entropy of a quantum system. This can be illustrated by considering the simple case of a state of $n=2$ qubits. The statevector 
\begin{align}
    \ket{\psi} = \sum_{i,\,j=0}^1 c_{ij}\ket{ij},~\text{with} \sum_{i,\,j=0}^1 |c_{ij}|^2=1,
\end{align}
can be expressed in the Schmidt decomposition as
\begin{align} \label{eq:schmidt}
    \ket{\psi} = \sum_{\alpha=1}^{\chi_s} \lambda_{\alpha}\ket{\xi_\alpha}_1\ket{\eta_\alpha}_2,
\end{align}
where $\chi_s$ is the \textit{Schmidt rank}, $\lambda_\alpha$ are the Schmidt coefficients, and $\{\ket{\xi_\alpha}_1\}, \{\ket{\eta_\alpha}_2\}$ are orthonormal bases in the space of the first and second qubit respectively. 
Using the decomposition~\eqref{eq:schmidt} in Eq.~\eqref{eq:ent_entropy}, the entanglement entropy between the two qubits then amounts to
\begin{align}
    S\qty( \rho_A ) = -\sum_{\alpha=1}^{\chi_s} \lambda_a^2\log\lambda_\alpha^2\, ,
\end{align}

In an MPS simulation we always have access to a subset of the Schmidt coefficients, since such representation is built by iteratively applying the Singular Value Decomposition (SVD), a procedure equivalent to Schmidt-decomposing a quantum state. The reason why we have access only to subsets of them is that we impose the following conditions on the Schmidt coefficients. Listing the coefficients in ascending order, i.e. $\lambda_0\geq \lambda_1 \geq\dots\geq \lambda_{\chi_s}$, then:
\begin{itemize}
    \item Schmidt coefficients whose ratio with $\lambda_0$ is smaller than $\epsilon$ are discarded. The value of $\epsilon$ in this work is fixed at $\epsilon=10^{-9}$;
    \item only the first largest $\chi_{max}$ coefficient are retained. The value $\chi_{max}$ is called maximum \textit{bond dimension}.
\end{itemize}
The approximation we are performing is the optimal one in terms of the represented entanglement. Then, the measure of entanglement for the MPS now becomes
\begin{align}
    S(\rho_A) = -\sum_{\alpha=1}^{\chi_{max}} \lambda_a^2\log\lambda_\alpha^2.
\end{align}
As explained in detail in Appendix~\ref{app:convergence}, despite the approximations, the faithfulness of the simulation can be easily monitored.
Finally, we remark that since we have constant access to the considered subset of Schmidt coefficients during the state evolution, we are able to compute the entanglement entropy of a quantum state without any computational overhead.

\subsection{Entanglement entropy in random quantum states}
\label{sec:HaarRandom}
In this section we briefly describe the entanglement features of uniformly distributed random pure quantum states, that is quantum states sampled according to the unique unitarily invariant probability distribution induced by the Haar measure. Denoting by $\mathbb{U}(n)$ the group of $2^n \times 2^n$ unitary matrices, there is a unique unitarily invariant probability measure $\mu(U)$ defined on the group, and such measure is called \textit{Haar measure}~\cite{HaydenAspectsGenericEntanglement2006, MeckesRandomMatrixTheory2019, EdelmanRandomMatrixTheory2005}. 
Unitary invariance corresponds to the requirement that the measure is invariant under translations in the space of unitary matrices, that is
\begin{equation*}
    \mu(M U) = \mu(U M) =  \mu(U) \quad U, M \in \mathbb{U}(n)\, .
\end{equation*}
The Haar measure induces a uniform probability distribution in the space of unitary matrices so that sampling a quantum state according to the Haar measure means randomly picking a state uniformly from the space of quantum states. We denote with $\mathcal{P}(n)$ such probability distribution.

We are interested in the entanglement features of random quantum states, particularly in the entanglement entropy. Let $\ket{\psi} \in (\mathbb{C}^2)^{\otimes{n}}$ be a quantum state of $n$ qubits sampled from the uniform distribution $\ket{\psi} \sim \mathcal{P}(n)$, and a bipartition of the $n$ qubits system in two subsystems $A$ and $B$, of size $n_A$ and $n_B = n - n_A$ respectively. Then, for $n_A \leq n_B$, the expected value of the entanglement entropy~\eqref{eq:ent_entropy} corresponding to this cut amounts to the Page value~\cite{HaydenAspectsGenericEntanglement2006, Page1993Entanglement}
\begin{equation}
\label{eq:haar_entanglement}
    \mathbb{E}[S(\rho_A)] = \sum_{j = d_B + 1}^{d_A d_B} \frac{1}{j} - \frac{d_A-1}{2d_{B}}\,,
\end{equation}
where $d_B = 2^{n_B}$, $d_A = 2^{n_A}$ are the local dimensions of the two subsystems, and the expectation value is over the uniform probability distribution $\mathbb{E}(\cdot) = \mathbb{E}_{\ket{\psi} \sim \mathcal{P}(n)}(\cdot)$. One can check that the entanglement is highest whenever the two partitions have equal size $n_A=n_B=n/2$ (for $n$ even, and similarly for $n$ odd, $n_A = {\lfloor\frac{n}{2}\rfloor}$ and $n_B = {\lceil\frac{n}{2}\rceil}$). 

From Eq.~\eqref{eq:haar_entanglement} it follows that $\mathbb{E}[S(\rho_A)] \geq \log d_A - d_A /2 d_B$~\cite{HaydenAspectsGenericEntanglement2006}, and since the maximum value of the entanglement entropy for such bi-partition is $\log d_A$, obtained if the subsystems $A$ and $B$ share maximal entanglement, one concludes that random states are generally highly entangled. Indeed, in ref.~\cite{HaydenAspectsGenericEntanglement2006} it was shown that the probability that a random pure state has entanglement entropy lower than $\log d_A - d_A /2 d_B$ is exponentially small. Thus, with very high probability, random quantum pure states are almost maximally entangled. 

\subsection{Quantum Neural Networks as Parameterized Quantum Circuits}\label{sec:QNN}
Currently available quantum devices are still too small and noisy to perform relevant fault-tolerant computations of notorious and efficient quantum algorithms, like Shor's factoring~\cite{NielsenChuang, Preskill2018NISQ}. For this reason, recent research has focused on a new paradigm of quantum computation based on so-called variational quantum algorithms (VQAs), which trade theoretical success guarantees with feasibility of execution, and are thought to be the most effective way to reach a quantum advantage in the near term, already with small quantum devices~\cite{CerezoPQC2021Review, Bharti2021NIQSReview, McClean2016VQAs}.

Variational quantum algorithms are based on PQCs, which are quantum circuits in which some of the unitary operations are characterized by \textit{variational} parameters to be adjusted in order to solve an optimization problem. The optimal parameters are found by minimizing a properly chosen cost (or loss) function encoding the task to be solved. Let $U_{\bm{\theta}}$ be the unitary evolution implemented by a quantum circuit with tunable parameters $\bm{\theta}$, and $O$ a Hermitian operator (an observable). The goal of variational quantum algorithms is to optimize the quantum circuit parameters $\bm{\theta}$ in order to minimize the expectation value (or variations thereof)
\begin{equation}
\label{eq:cost_function}
    f(\bm{\theta}) = \expval{\mathcal{O}}_{\bm{\theta}} = \Tr[O\, U_{\bm{\theta}} \rho U^\dagger_{\bm{\theta}}]
\end{equation}
where $\rho$ is an initial quantum state, generally set to the ground state $\rho = \dyad{\bm{0}}$. This is achieved by means of an iterative hybrid quantum-classical approach where the quantum computer is used to estimate the cost function~\eqref{eq:cost_function}, and given such value, the classical computer proposes new variational parameters according to an optimization method, the most common one being gradient descent. 

There is freedom in the choice of the gate sequence defining the parameterized unitary $U_{\bm{\theta}}$, and a choice of its structure is referred to as variational \textit{ansatz}. For example, the unitary could be composed of a layer of Pauli rotations around the $X$-axis on each qubit $R(\theta) = \exp(-i\theta X/2)$, followed by a layer of \textsc{Cnot}s acting on pairs of neighboring qubits. This is in fact the general blueprint of variational quantum circuits, as they are generally created by repeating single-qubits parameterized rotations followed by multi-qubits operations which introduce entanglement into the computation. Examples of parameterized quantum circuits are shown in Fig.~\ref{fig:main_figure}.

\paragraph{Quantum Neural Networks.} As it is often the case with learning tasks, either classical or quantum, the goal is to solve a problem given access to a dataset of inputs $\mathcal{X} = \{\bm{x}_i\}_i$, representative of the task to be solved. Whenever data is involved, variational quantum circuits are often referred to as Quantum Neural Networks, as they share the very same idea as their classical counterpart: learn patterns in input data by adjusting tunable parameters through the iterative minimization process known as \textit{training}.
In this case, the quantum circuit of the neural network depends on two sets of parameters $\bm{x}$ and $\bm{\theta}$, the former being the input data to be analyzed, and the latter the variational parameters to be adjusted (i.e. the \textit{weights} of the neural network). In the quantum machine learning jargon, the encoding scheme used to load the input data onto the quantum computer is known as \textit{feature map}, and consists of a unitary operation parameterized by $\bm{x}$. We will denote such feature encoding gate with $\mathcal{F}(\bm{x})$, where $\bm{x}\in \mathcal{X}$. As with the variational unitary, there is no standard choice for a feature map, and one has to pick a specific ansatz, ideally biasing the choice towards architectures built using knowledge of the problem to be solved~\cite{SkolikEquivariant, MeyerSymmetries}. Summing up, a general QNN can be then expressed as
\begin{equation}
\label{eq:qnn_equation}
\begin{aligned}
    U_{\text{QNN}}(\bm{x};\bm{\theta})  & = \prod_{i=L}^1 V(\bm{\theta}_i)\mathcal{F}(\bm{x})\\
    & =  V(\bm{\theta}_L)\mathcal{F}(\bm{x})\cdots V(\bm{\theta}_1)\mathcal{F}(\bm{x})\,, 
\end{aligned}
\end{equation}

where $\mathcal{F}(\bm{x})$ is the feature map ansatz depending on the input data $\bm{x}$; $V(\bm{\theta}_i)$ is a variational ansatz depending on trainable parameters $\bm{\theta}_i\in \bm{\theta} = (\bm{\theta}_1, \cdots, \bm{\theta}_L)$ with $\bm{\theta}_i\in \mathbb{R}^p$; and $L$ is the number of repetitions (also referred to as \textit{layers}) of the such layered structure. 

It was recently shown that uploading the input data multiple times throughout the circuit is essential for quantum neural networks to model higher-order functions of the inputs~\cite{Schuld2020Encoding, Theis2020Expressivity}. Such procedure is now standard practice in the quantum neural network-based quantum machine learning, and it is called \textit{data reuploading}~\cite{Torrontegui2019Perceptron, Perez2020Reuploading}. Notice that the input data in the feature map in Eq.~\eqref{eq:qnn_equation} is the same in every layer, while the variational blocks $V$ use a different parameter vector in every layer. In Fig.~\ref{fig:main_figure}(a) we give a graphical representation of the general structure of QNNs. As for the explicit implementation of $\fmap$ and $\varans$, there is no fixed choice and these are usually composed of single qubit rotations followed by entangling operations, either fixed (e.g. \textsc{Cnot}s) or themselves parameterized (e.g. controlled rotations). See Fig.~\ref{fig:main_figure}(c) for some prototypical examples of parameterized blocks proposed in the literature~\cite{AbbasPowerQNN2021, SimPQCs2019}, which we will consider throughout the manuscript.

\subsection{Randomness, Entanglement and Trainability. \label{par:2des_ent_bp}}
One of the hardest theoretical challenges affecting quantum machine learning models is the emergence of so-called \textit{barren plateaus} (BPs) in their optimisation landscape~\cite{McCleanBarren2018}. BPs are regions in parameter space where the loss function is essentially flat, with no interesting minimising direction, so that it is not possible to train the model efficiently and independently of the optimization methods used, be it gradient-based~\cite{Schuld2019Gradients} or gradient-free~\cite{Arrasmith2020effect}. Different sources can lead to the unfolding of barren plateaus, and these can be broadly grouped into three main categories: randomness-induced BP~\cite{McCleanBarren2018, Holmes2021connecting, MarreroBarrenEntanglement2021}, BP induced by \textit{global} cost functions defined with observables having support on a large number of qubits~\cite{CerezoBarrenLocalCost2021}, and eventually noise-induced BP~\cite{WangNoiseinducedBarrenPlateaus2021}.

In this work we are concerned with the former type of barren plateaus, that roughly occur when parameterized quantum circuits, when initialised with random parameters, resemble general random unitaries. Indeed, despite being quite limited in terms of qubits connectivity and gate operations, common instances of parameterized quantum circuits are often found to behave as unitary 2-designs, that is they efficiently approximate the statistics of Haar-random unitaries up to the second moment~\cite{Dankert2des2009}. In this case, then one can prove that the variance of the gradients of any cost function $f(\bm{\theta})$ defined on the circuit will vanish exponentially with the number of qubits $n$, namely~\cite{Holmes2021connecting}
\begin{equation}
    \label{eq:bp_vanishing}
    \text{Var}_{\bm{\theta}}[\partial_k f(\bm{\theta})] \in \order{b^{-n}}\, \quad \new{b>1}\,,
\end{equation}
where $f(\bm{\theta})$ is as in Eq.~\eqref{eq:cost_function}. Specifically, the cost function concentrates around its mean value and stays constant almost everywhere in parameter space~\cite{Arrasmith2021equivalence}, which makes training unfeasible. 

Vanishing gradients are used as a \textit{witness} to assess whether a parameterized quantum circuit resembles a unitary 2-designs. Of course, this is only \textit{necessary} but not \textit{sufficient} condition, as one can easily devise a circuit that is not a 2-design but has vanishing gradients, for example using a global cost with a shallow  circuit~\cite{CerezoBarrenLocalCost2021}.

In addition to vanishing gradients, another witness of randomness is the entanglement generated inside the circuit~\cite{MarreroBarrenEntanglement2021}. Indeed, as discussed previously in Sec.~\ref{sec:HaarRandom}, random quantum states are almost maximally entangled, so one can use the maximality of entanglement generated by a parameterized circuit as an indicator of the resemblance to a random unitary evolution. As for vanishing gradients, the presence of large entanglement is however only a necessary but not sufficient condition for randomness, as a simple shallow circuit composed of Hadamards and \textsc{Cnot}s can create maximally entangled states (GHZ states), which are clearly not random. As discussed in~\cite{SackBPShadows2022}, the so-called entanglement-induced BPs~\cite{MarreroBarrenEntanglement2021, Patti2021EntDevisedBP} provide an alternative yet equivalent description of local cost barren plateaus (circuits with global costs always suffer of vanishing gradients~\cite{CerezoBarrenLocalCost2021}, regardless of randomness), as they both stem from the proximity of parameterized quantum circuits to unitary 2-designs. 

Indeed, if a circuit is a unitary 2-design, then the average entanglement entropy of any subsystem $A$ of dimension $d_A$ ($d_A \leq d_B$) will be already very close to its maximal value~\cite{SackBPShadows2022, LiuEntanglementQuantumRandomness2018}
\begin{equation}
\label{eq:2des_vonn}
    \log d_A - 1 \leq \mathbb{E}_{\bm{\theta}}[S(\rho_A)] \leq \log d_A\,,
\end{equation}
and approaches the Page value~\eqref{eq:haar_entanglement} for truly Haar-random states. We provide a proof of Eq.~\eqref{eq:2des_vonn} based on the Rényi 2-entropy in  Appendix~\ref{app:reny}.

Recent investigations quantify the tight connection between trainability and randomness in terms of the \textit{expressibility}, roughly defined as the ability of a parameterized quantum circuit to address the full unitary space~\cite{SimPQCs2019}, and show that highly expressible ansätze have flatter loss landscape, hence they are harder to train~\cite{Holmes2021connecting}. We further discuss the expressibility measure in Sec.~\ref{sec:Discussion}.

To summarise, while the presence of entanglement is a necessary ingredient to avoid classical simulability, its uncontrolled growth is likely to signal the emergence of barren plateaus. The evaluation of the entangling capabilities of parameterized quantum circuits is then a valuable diagnostic tool to provide information both on the classical simulability and trainability issues of quantum machine learning models. At last, we note that although various methods have been put forward to mitigate the occurrence of BPs~\cite{Grant2019Initialization, Volkoff2020Correlating, Skolik2020Layerwise}, including proposals based on entanglement control~\cite{SackBPShadows2022, Patti2021EntDevisedBP, Joonho2021EntBP}, these remain a bottleneck for scaling up quantum machine learning computations based on variational circuits. 

\section{Results}\label{sec:Results}
We now proceed to analyze the entanglement production in various quantum neural network architectures with different feature maps and variational ansatz, obtained composing the circuit blocks shown in Fig.~\ref{fig:main_figure}. In particular, we take as a prototypical example the QNN introduced in \cite{AbbasPowerQNN2021}, argued as a good candidate for quantum machine learning applications in terms of capacity and expressibility, possibly achieving an advantage over classical counterparts. Such QNN model uses as feature-map $\mathcal{F}(\bm{x})$ the so-called $\textsc{ZZFeatureMap}$ firstly introduced in~\cite{Havlicek2019QSVM} as a classically-hard map to load classical data on a quantum state in a non-linear fashion. The variational block $V(\bm{\theta})$ is instead composed of single qubit rotations followed by entangling operations. In order to better understand the effect of every single operation in the quantum circuit, we also consider variations of the QNN introduced above, varying both the feature map, the variational form, and the entangling topology. All considered circuit blocks are graphically represented in Fig.~\ref{fig:main_figure}. 

Be $U_L(\bm{x}, \bm{\theta})$ the unitary representing a specific quantum neural network with $L$ layers with input data $\bm{x}=(x_1, \hdots, x_m) \in \mathbb{R}^m$, and variational parameters $\bm{\theta}=(\theta_1, \hdots, \theta_p) \in \mathbb{R}^p$, see Eq.~\eqref{eq:qnn_equation}. We consider random instances of such QNN by sampling both the inputs and the variational parameters according to the uniform distribution $x_i, \theta_i \sim \text{Unif}(0, \pi)$, hence obtaining a collection of QNNs $U_L = \{U_L(\bm{x}_i, \bm{\theta}_i)\,, i=1, \hdots, M\}$. Then, we study the entanglement entropy properties of each of these instances and average the result over the $M$ trials (unless stated otherwise, we take $M=100$). Thus, when in the following we refer to the entanglement entropy of a quantum circuit, we are always denoting the average over $M$ realizations of that circuit. In order to evaluate the influence of the depth on the entanglement, we repeat this analysis by increasing the number of layers in the quantum neural network $L=1,\hdots, L_{max}$. 

Note that although the total number of parameters (inputs and parameters) depends on the specific feature map and variational form used, for the considered circuits such difference generally amounts to a constant and does not have a relevant impact on the results. In Tab.~\ref{tab:summary_circuit} we report the number of parameters in each circuit template analyzed in this work. We anticipate that while the number of parameters in the considered quantum circuits only scales polynomially with the system size $n$, these are found to be sufficient to reproduce some entanglement features of random unitaries, which are instead characterized by an exponential number parameters. This is in agreement with results on random quantum circuits that states that polynomial resources are sufficient to approximate unitary designs~\cite{harrowRandomQuantumCircuits2009, HaferkampRandom}. We refer to Sec.~\ref{sec:Discussion} for an extended discussion.

\begin{table*}[ht]
    \centering
    \begin{tabular}{ccccc} \toprule
                                      & Abbr. &\multicolumn{3}{c}{Number of parameters}\\
                                      &       & \textsc{Linear} & \textsc{Circular} & \textsc{Full} \\ \midrule
        \textsc{Circuit 1}            & $\circuit{1}$   & $2n$  &   $2n$  & $2n$      \\     
        \textsc{Circuit 2}            & $\circuit{2}$   & $n$   & $n$  & $n$      \\
        \textsc{Circuit 3}            & $\circuit{3}$   & $2n-1$& $2n$ & $\frac{n^2+n}{2}$ \\
        \textsc{Circuit ZZFeatureMap} & $\circuit{zz}$  & $n$   & $n$ & $n$       \\
    \bottomrule
    \end{tabular}
    \caption{Number of parameters for each considered circuit and their relative entanglement topology. Notice that, while the number of parameters remains constant for  $\circuit{zz}$ as shown in the table, the number of parametric gates varies analogously to $\circuit{3}$.}
    \label{tab:summary_circuit}
\end{table*}

\subsection{Alternating vs.~Sequential data reuploading}
\begin{figure}[ht]
    \centering
    \includegraphics{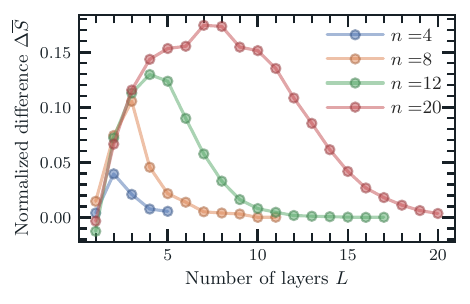}
    \caption{Normalized entanglement difference $\Delta \overline{S}$, as defined in Eq.~\eqref{eq:ent_diff} for different numbers of qubits. The used QNN is defined by $\fmap = \circuit{zz}$ and $\varans = \circuit{2}$. All the data points are obtained by averaging over $10^3$ realizations.}
    \label{fig:ent_diff}
\end{figure}

As a first analysis, we study the difference in entanglement growth between a standard QNN using an \textit{alternated} repetitions of feature maps and variational forms (as in Fig.~\ref{fig:main_figure}), and one in which we have first $L$ repetitions of the feature map followed by $L$ repetitions of the variational form. We call this structure \textit{sequential}. The former leverages an alternated evolution of the quantum state which is typical of quantum neural networks using a data reuploading scheme~\cite{Perez2020Reuploading, Schuld2020Encoding, Theis2020Expressivity}. The latter instead uses an initial data-dependent evolution followed by a trainable unitary, thereby creating an architecture similar to quantum kernel machines~\cite{SchuldKernel}. While the two structures (alternated and sequential) may be mapped to each other using ancillary qubits~\cite{JerbiQMLKernel2021}, they can have rather different performances, and we hereby show how they also create entanglement in a different way. Specifically, given the two unitary evolutions, namely the fixed input-dependent feature map $\mathcal{F}(\bm{x})$ and the varying parameterized variational form $V(\bm{\theta}_i)$, one expects the alternated dynamics
\[
U_{alt} = V(\bm{\theta}_L)\mathcal{F}(\bm{x})\cdots V(\bm{\theta}_1)\mathcal{F}(\bm{x})\,,
\]
to introduce randomness at a faster rate than the sequential process
\[
U_{seq} = \prod_{i=1}^L V(\bm{\theta}_i)\,\prod_{i=1}^L\mathcal{F}(\bm{x})\,,
\]
and hence introduce more entanglement in the system. Such intuition is confirmed by the numerical results, and may be understood as a consequence of the universality of the alternating dynamics proved for example for QAOA circuits~\cite{LloydUniversalityQAOA, MoralesUniversalityQuantumApproximate2020}. 

Here we use $\mathcal{F}=\circuit{zz}$ and $V=\circuit{2}$, as defined in Fig.~\ref{fig:main_figure}, both with linear topology. Be $S^{alt}$ and $S^{seq}$ the entanglement entropy of the bipartition with an equal number of qubits, which is generally the highest, for the alternating and sequential structure, respectively. We define the normalized difference as 
\begin{align}
    \label{eq:ent_diff}
    \Delta \overline{S} = \frac{S^{alt}-S^{seq} }{ (S^{alt}+S^{seq})/2}\,,
\end{align}
and study its behavior as the depth of the quantum circuit is increased, as shown in Fig.~\ref{fig:ent_diff}. 

The metric is always positive and features a maximum, implying that the alternated structure is creating entanglement \textit{faster} (i.e. with fewer layers) than its non-alternated counterpart. Note that for $L=1$ layers the two structures are identical, so the generated entanglement is the same up to the statistical error, which explains why all the curves start around zero. At a high number of repetitions, the two structures tend to the same value, showing a $\Delta\overline{S}\simeq 0$, which can be understood in light of the results presented in the following sections: as the number of layers of a QNN is increased, the entanglement rapidly converges to that of a Haar-distributed random state, thus the alternated and non-alternated structure eventually converge to the same value. Given the higher entanglement production rate of the alternated structure, in the following analysis, we shall focus on this structure only.

\subsection{Entanglement distribution across bonds}\label{sec:EntanglementBond}
\begin{figure}[!ht]
    \centering
    \includegraphics{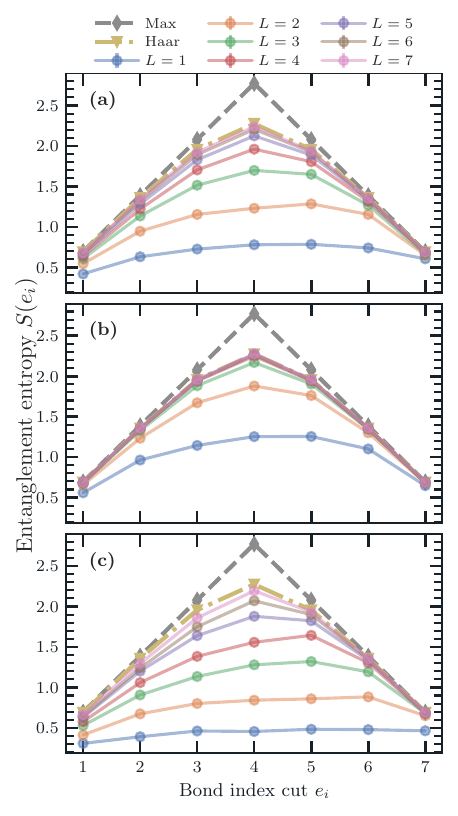}
    \caption{Average entanglement entropy across bonds for a system composed of $n=8$ qubits, where $e_i$ is the bond connecting qubit $i$ and $i+1$, as in Fig.~\ref{fig:main_figure}. The curves represent different numbers of layers $L$ in the quantum neural network. \textbf{(a)} QNN with structure: $\mathcal{F} = \circuit{zz}$, $V = \circuit{2}$, both with linear entanglement. \textbf{(b)} QNN with structure: same as in (a) but with circular entanglement. \textbf{(c)} QNN with structure: $\mathcal{F} = \circuit{1}$ which has no entangling gates, and $V = \circuit{2}$ with linear entanglement.}
    \label{fig:bond_entanglement}
\end{figure}
It is natural to ask how the choice of the feature map, the variational form, and the entangling topology impact the growth of entanglement of the quantum state. In this section, we start to explore this question by studying how entanglement is distributed across all possible ordered bi-partitions of the $n$ qubits in the network. That is, given an MPS representation as in Fig.~\ref{fig:main_figure}(b), we study the entanglement entropy corresponding to each bond in the linear chain. Denoting with $e_i$ the bond connecting qubit $q_i$ and $q_{i+1}$, the entanglement entropy of that bond is (see Eq.~\eqref{eq:ent_entropy})
\begin{equation}
\label{eq:ent_entropy_bonds}
\begin{aligned}
    S(e_{i}) & = -\Tr[\rho_{[1:i]}\log \rho_{[1:i]}] \\
    \rho_{[1:i]} & = \Tr_{i+1, \hdots, n}[\rho]
\end{aligned}\,\,,
\end{equation}
where $\rho_{[1:i]}$ is the reduced density matrix of all the qubits up to the $i$-th one, and $\rho$ is the state obtained from the quantum neural network $\rho = U_L(\bm{x}, \bm{\theta}) \dyad{\bm{0}} U_L(\bm{x}, \bm{\theta})^\dagger$.

In Fig.~\ref{fig:bond_entanglement} we show the entanglement entropy distribution for the case of $n=8$ qubits using three different quantum neural networks architectures: in panel (a) the one proposed in~\cite{AbbasPowerQNN2021} with feature map $\mathcal{F}=\circuit{zz}$, variational ansatz $V=\circuit{2}$, both with linear entanglement; in (b) same as before but using a circular entanglement topology; and eventually in panel (c) a simpler circuit using a tensor product feature map $\mathcal{F}=\circuit{1}$ which encodes data independently on each qubit, followed by the same variational ansatz $V=\circuit{2}$, again with linear entanglement both. For reference, it is also shown the expectation value of the entanglement entropy for Haar-random quantum states evaluated with Eq.~\eqref{eq:haar_entanglement}, as well as an upper bound given by the highest possible entanglement $\log(\text{min}(d_A, d_B))$, obtained if the two partitions $A$ and $B$ were maximally entangled. Note that while we report only the simulation data for $n=8$, the discussion has general validity as identical results hold for all tested numbers of qubits, $n=2,\,\hdots,\,20$.

First of all, the findings agree with the intuition that deeper circuits are able to create higher entangled states with respect to shallower ones, in accordance with results from~\cite{SimPQCs2019}. In particular, the entanglement entropy is higher at the center of the chain. 
Clearly, depending on the specifics of the QNN, the entanglement grows faster in certain architectures with respect to others. Regarding the effect of the entangling topology, comparing panels (a) and (b) we see that circular connections produce greater entanglement compared to the nearest-neighbors interaction and that such entanglement grows at a faster rate as the number of layers is increased. As for the choice of the feature map, since the QNN in panel (c) produces entanglement only through the entangling gate in the variational blocks, its entanglement is lower and also grows slower with respect to the QNN in panel (a), even though it has twice the number of parameters in the feature map.

Interestingly, however, as the number of layers approaches the number of qubits $L\approx n$, all investigated QNNs converge to the same values, that is those obtained for random states sampled from the uniform Haar distribution. Deep enough QNNs are then flexible enough to reproduce the same entanglement spectrum of a random state, which, as discussed in section~\ref{sec:HaarRandom}, are very highly entangled. Again, even though the measure of entanglement is different, this is in agreement with the results presented in~\cite{SimPQCs2019}, where the convergence to the Haar distribution is encountered for various parameterized quantum circuits, and also with other results in the literature regarding the properties of random quantum circuits to approximate the Haar distribution~\cite{BrandaoLocalRandomQuantum2016, Harrow2018approximate}. We will discuss this more in detail in Sec.~\ref{sec:Discussion}. 
A more in-depth analysis of the convergence is the subject of the next section.

\subsection{Entanglement scaling with increasing depth}\label{sec:EntanglementScaling}
\begin{figure*}[ht]
    \centering
    \includegraphics[width=\textwidth]{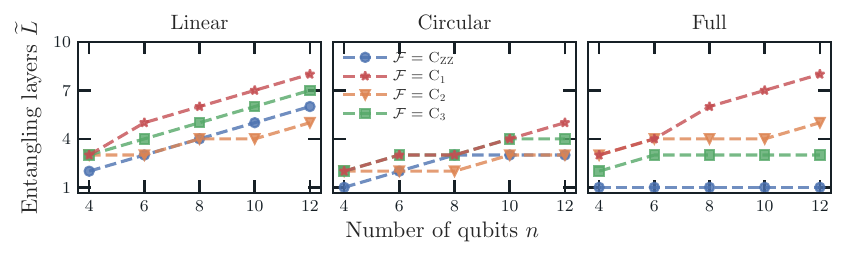}
    \caption{Entangling layers $\widetilde{L}$, i.e number of layers to reach $90\%$ of the Haar entanglement, versus the number of qubits. The analysis is carried out over four different QNN architectures, each evaluated with different entangling topologies (\textit{linear}, \textit{circular}, and \textit{full}). The architectures leverage the same variational form $\varans$, while the feature map $\fmap$ is changed, as reported in the legend. }
    \label{fig:ent_scaling}
\end{figure*}

In order to better understand the entanglement scaling properties of QNNs, we introduce a new quantity, defined as the total entanglement entropy $S_{tot}$ created in the MPS chain
\begin{equation}
\label{eq:total_entanglement}
    S_{tot} = \sum_{i=1}^{n-1} S(e_i)\, ,
\end{equation}
which is the sum of the entanglement entropy of all the ordered bipartitions of the quantum state. We use this global measure to quantify how fast QNNs approach the Haar distribution in terms of overall entanglement production. In particular, we define a new figure of merit, the entangling layers $\widetilde{L}$, defined as the number of layers needed by an architecture to reach $90\%$ of the total entanglement of a Haar distributed state $S_{tot}^{\text{Haar}}$, namely
\begin{align}
\label{eq:tildeL}
    \widetilde{L} = \text{min \#} \text{ of layers } \; \text{s.t.} \;\;\; S_{tot} \geq  0.9\,S_{tot}^{\text{Haar}}.
\end{align}
The choice of the $90\%$ threshold allows to select states that are already very close to the Haar-random value, and avoids undesired oscillating behaviours obtained when higher thresholds are used, e.g. $99\%$, which are caused by statistical fluctuations (recall that every QNN is sampled multiple times with different parameters to calculate averages).

In Fig.~\ref{fig:ent_scaling} we show the behavior of $\widetilde{L}$ for four different QNNs as the number of qubits is increased. Note that each QNN is considered with all the three possible entangling topologies (\textit{linear}, \textit{circular} and \textit{full} as defined in Fig.~\ref{fig:main_figure}). At last, note that all QNNs leverage the same variational form $\varans = \circuit{2}$, while the feature map is changed, as reported in the legend. 

First, we observe that the entangling layers display a linear behavior when a linear entanglement topology is used. This means that the number of layers needed to entangle the system scales linearly with the size of the system. The behavior changes abruptly when we move to a circular or full entangling topology. All architectures display a faster entanglement production when passing from a linear to a circular topology, as can be seen from the lower slope of the curves. The all-to-all connectivity speeds up entanglement production only for $\fmap = \circuit{zz},\,\circuit{3}$, while the circuits $\fmap = \circuit{2},\,\circuit{1}$ show essentially the same behavior of the linear case. We now proceed to discuss more in detail such results.

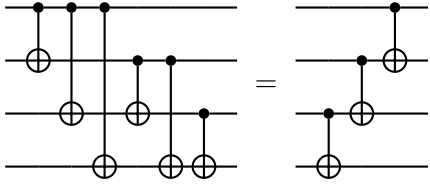
\begin{figure}[!ht]
    \centering
    \[
    \begin{quantikz}[column sep={12.4pt,between origins}, row sep={20pt,between origins}]
    & \ctrl{1} & \ctrl{2} & \ctrl{3} 	& \qw  		& \qw 	 & \qw		& \qw \\
    & \targ{}  & \qw		& \qw 			& \ctrl{1} & \ctrl{2} & \qw		& \qw \\ 
    & \qw 		& \targ{}  & \qw			& \targ{} 	& \qw		 & \ctrl{1}& \qw \\
    & \qw 		& \qw 		& \targ{}		 & \qw 		& \targ{} & \targ{}	& \qw 
    \end{quantikz}
    =
    \begin{quantikz}[column sep={12.4pt,between origins}, row sep={20pt,between origins}]
    & \qw 	 	& \qw		& \ctrl{1}&\qw \\
    & \qw 		& 	\ctrl{1}	& \targ{} & \qw \\ 
    & \ctrl{1}	& \targ{} & \qw		& \qw \\
    & \targ{} & \qw		& \qw 		& \qw 
    \end{quantikz}
    \]
    \caption{Circuital identity between a full entangling map made of only \textsc{Cnot}s and the adjoint of a linear entangling map.}
    \label{fig:cnot_equivalence}
\end{figure}

We start comparing the entangling capabilities of $\circuit{zz}$ vs. $\circuit{2}$. Both with linear and circular entangling topology, $\circuit{2}$ is able to produce entanglement essentially at the same rate as $\circuit{zz}$, despite $\circuit{2}$ being of a much simpler structure, with half the number of two-qubit gates. However, things change dramatically using a full entangling map, as the QNN reaches the $90\%$ threshold already at $\widetilde{L}=1$, while $\circuit{2}$ needs more layers, showing the same dependence of a linear connectivity. While counter-intuitive at first, is it easy to see that the entanglement generated by $\circuit{2}$ with a \textit{full} architecture is indeed equivalent to the \textit{linear} one. This is due to a simple circuit identity regarding networks of \textsc{Cnot}s reported in Fig.~\ref{fig:cnot_equivalence}. Such circuital identity holds for any number of qubits, which makes the full entangling map as shown in Fig.~\ref{fig:main_figure} just as a linear entangling map in disguise (in particular, it is the inverse of the linear entangling map). See Appendix~\ref{app:entangling_maps} for a more precise statement, discussion and proof. Such circuital identity thus explains the equivalence of the yellow ($\fmap = \circuit{2}$) and red ($\fmap = \circuit{1})$ curves between the first and last plot of Fig.~\ref{fig:ent_scaling}.

Such equivalence clearly does not hold if controlled rotations are used instead of \textsc{Cnot}s. Indeed, the feature map $\fmap = \circuit{3}$ uses controlled rotations with independent random parameters, and given that these gates do not cancel out, the entanglement is always increasing going from low to high connectivity. Note that such increase is mainly due to the feature map, as the variational ansatz $\varans = \circuit{2}$ is the same as other structures, suffering from the \textsc{Cnot}s cancellation issue described above. 

For comparison, we also show the performances of a QNN with the tensor product feature encoding $\fmap =  \circuit{1}$, using no entangling operations. Interestingly, even if this QNN uses two-qubit interactions only inside the variational blocks, these are sufficient to create entanglement similar to other considered QNNs, even at a slower yet comparable rate.

We report in Appendix \ref{app:ext_ent_scaling} the complete simulation results detailing the evolution of the entanglement with the depth of the circuit, for different numbers of qubits.

\subsection{Entanglement Speed}\label{sec:MPSSimulation}
So far we have presented numerical evidence for the entanglement production in QNNs up to a maximum of 20 qubits. In the following we extend the analysis leveraging MPS to simulate quantum systems of bigger size up to 50 qubits, with a maximum bond dimension of $\chi_{\max}=4096$. More importantly, we show how the entanglement growth follows a behavior that is specific to each particular QNN architecture and the number of layers considered, but independent of the number of qubits in the circuit. We can thus uniquely assign an \textit{entanglement speed} value to each QNN, which, we stress again, only depends on the choice of the ansatz, and holds identically for any instantiation of that QNN with arbitrary number of qubits.

Taking into account the entanglement growth discussed in Sec.~\ref{sec:EntanglementScaling}, we restrict the analysis to a linear architecture, to increase as much as possible the number of layers we can correctly simulate with tensor networks techniques. Indeed, the entanglement production with a circular or full topology is too fast to allow for a convergent simulation with MPS for deep circuits.

Furthermore, we introduce the maximum Haar entanglement entropy, defined as the maximum across all bond entropies for a given number of qubits, as
\begin{equation}
\begin{aligned}
\label{eq:haar_max}
    S^{\text{Haar}}_{n, \max}&=\max_A\big( \mathbb{E}[S(\rho_A)] \big)\\ &\approx \frac{n}{2}\log 2 - \frac{1}{2}\quad \text{for}~n_A=\frac{n}{2} \gg 1,
\end{aligned}
\end{equation}
where the approximation in the second line has an errors that scales as $\order{2^{-n/2}}$, see Appendix~\ref{app:haar} for its derivation. Thus, for $n\geq30$ qubits, when the exact computation of the Haar entanglement entropy is unfeasible, we employ the approximated Eq.~\eqref{eq:haar_max}. Finally, we define the normalized entanglement entropy $\widetilde{S}_n$ as:
\begin{align}
\label{eq:normalised_entanglement}
    \widetilde{S}_{n}=\frac{\max_{e_i}\left[S(e_i)\right]}{ S^{\text{Haar}}_{n, \max}}\,.
\end{align}
We stress that $\widetilde{S}_n$ is normalized to the maximum Haar entanglement for a fixed $n$,  not to the real maximum of the entanglement, which would be $S=\frac{n}{2}\log 2$ for the equal size bipartition. 

\begin{figure}[t]
    \centering
    \includegraphics{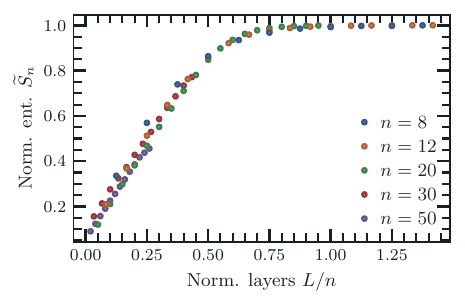}
    \caption{Normalized entanglement $\widetilde{S}_n$~\eqref{eq:normalised_entanglement} versus the normalized number of layers $L/n$, for different number of qubits $n$, and for the QNN defined by $\fmap=\circuit{zz},~\varans=\circuit{2}$ with linear connectivity. All the normalized entanglement points follow the same curve, independently of the system size $n$. The points for $n=8,\, 12,\, 16,\, 20$ are obtained by averaging over $10^3$ samples, while for $n=30,\, 50$ the averages are over 10 samples.}
    \label{fig:normalized_entanglement}
\end{figure}

In Fig.~\ref{fig:normalized_entanglement} we show the evolution of $\widetilde{S}_{n}$ versus the normalized number of layers $L/n$ for $n \in \{8,\, 12,\, 16,\, 20,\, 30,\, 50\}$ qubits, for the QNN defined with $\fmap=\circuit{zz}, \; \varans=\circuit{2}$ with linear connectivity. We note that all the points, independently of the system size $n$, follow the same curve: an initial linear growth of the entanglement is followed by a saturation to the Haar-random value for the entanglement entropy~\eqref{eq:haar_entanglement}. In particular, we check this behaviour also at large system sizes with $n=30,\, 50$ qubits and circuits with up to $L=11$ layers, and confirm that such scaling is indeed size independent. See Appendix~\ref{app:convergence} for a discussion on the errors introduced by truncation in the MPS representation for simulations with $n=30,\, 50$ qubits.

\begin{table*}[t]
    \centering
    \begin{tabular}{ccc} \toprule
         Feature map $\fmap$      & Variational ansatz $\varans$ & Entangling speed $v_s$ \\ \midrule
         $\circuit{zz}$ & $\circuit{2}$   & $(1.8\pm0.1)$  \\ 
         $\circuit{zz}$  & $\circuit{3}$ & $(0.59\pm0.02)$\\ 
         $\circuit{1}$  & $\circuit{3}$   & $(0.316\pm0.006)$  \\ \bottomrule
    \end{tabular}
    \caption{Entangling speed, i.e. a measure of how fast the entanglement is created, for different QNN architectures. These results are obtained using up to $16$ qubits.}
    \label{tab:ent_speed}
\end{table*}

Inspired by the behavior of $\widetilde{S}_n$, we introduce a measure for the entanglement production which is specific to a given QNN architecture (feature map plus variational ansatz) and independent of the number of qubits.
Borrowing from the literature on random quantum circuits, it is known that the entanglement of a system undergoing random evolution initially grows linearly in time (depth of the circuit) before reaching the plateau of Haar random states~\cite{Calabrese2005EntEntropyGrowth, TianciEntGrowth2019, NahumEntGrowth2017, Joonho2021EntBP}. Indeed, as clear from Fig.~\ref{fig:normalized_entanglement}, we observe the same initial linear growth, and thus we define the \textit{entangling speed} $v_s$ as
\begin{equation}\label{eq:ent_speed}
    \widetilde{S}_n \propto v_s \cdot \bigg(\frac{L}{n} \bigg)\, \text{ for } \widetilde{S}_n\leq 0.5,
\end{equation}
where $0.5$ is a threshold such that the linear behavior holds. 
The entangling speed can thus be obtained by fitting the curve in Fig.~\ref{fig:normalized_entanglement} with the linear function~\eqref{eq:ent_speed} in the appropriate range. We report in Tab.~\ref{tab:ent_speed} the entangling speed for a subset of the inspected architectures, and notice that entanglement is produced at sensibly different rates. In agreement with the findings of Sec.~\ref{sec:EntanglementScaling}, we see that for a linear topology the circuit $\circuit{2}$ builds the entanglement at the fastest rate. Indeed, fixing the feature map to $\fmap=\circuit{zz}$, $\circuit{2}$ produces entanglement $3$ times faster than $\circuit{3}$.

\begin{figure}
    \centering
    \includegraphics{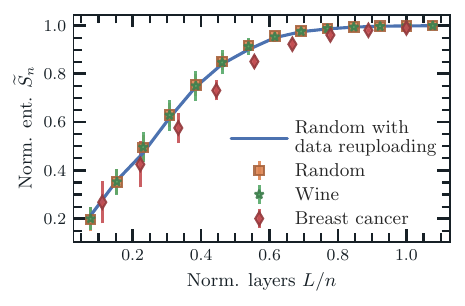}
    \caption{\new{Normalized entanglement using three different ways to sample the inputs $\bm{x}$ in the data encoding blocks. (\textit{i}) ``Random with data reuploading'' indicates the case of random synthetic inputs with data reuploading, as shown also in Fig.~\ref{fig:normalized_entanglement}. (\textit{ii}) ``Random'' indicates the case of random synthetic inputs without data reuploading, that is encoding blocks in different layers have different random parameters. (\textit{iii}) ``Wine'' and ``Breast cancer'' indicates inputs drawn from the corresponding real-world dataset, used with data reuploading. In all cases, the parameters in the variational blocks are sampled from the uniform distribution $\text{Unif}(0,\, \pi)$. Such distribution is used also to sample the synthetic random inputs. Data points for real-world datasets are obtained by first averaging over $10$ realization for each sample in the dataset, and then averaging again over the whole dataset. Results for random inputs without data reuploading (yellow square markers) are obtained by averaging over $2000$ realizations of the circuits. The error bars show the standard deviation of the mean. Error bars associated with random inputs with data reuploading (blue curve) are not shown to avoid cluttering but are of comparable size with the other points.}}
    \label{fig:datasets}
\end{figure}

To further characterise the applicability of the entangling speed, we show that the behavior of Fig.~\ref{fig:normalized_entanglement} evaluated for random circuits also holds when the input data $\bm{x} \in \mathbb{R}^n$ in the feature map $\mathcal{F}(\bm{x})$ are not drawn from the uniform distribution, but rather from real-world datasets. In particular, we select two common datasets in the machine learning literature, the wine~\cite{wine_dataset} and breast cancer~\cite{breast_cancer_dataset} datasets, and calculate the entanglement generated in the circuit when these data are fed into the feature maps (variational blocks are still populated with random parameters as before). The results presented in Fig.~\ref{fig:datasets} are obtained by rescaling all the features of the datasets in the interval $[0, \pi]$. \new{For each sample in the dataset, we average over $M=10$ runs with randomly drawn parameters for the variational ansatz. The results shown in the figure are then obtained as the average over the whole dataset.} \old{and then averaging over all data points in the datasets.} The wine dataset ($n=13$ features, hence $n=13$ qubits, \new{and $178$ samples}) follows perfectly the theoretical curve, and the breast cancer ($n=9$ features, hence $n=9$ qubits, \new{and $286$ samples}) only slightly deviates from it, producing entanglement at a smaller rate. We then conclude that the entangling speed depends primarily on the architecture of the circuit rather than the actual values of the parameters. Clearly, this holds for reasonably distributed data features, that is excluding pathological cases of values being either zero or concentrating around it.
\new{Finally, to verify that the QNN architecture is ultimately responsible for the entanglement speed, we analyze random circuits where the encoding blocks do not share the parameters, but these are sampled independently for each layer, thus effectively removing the data-reuploading feature. This case is portrayed in Figure~\ref{fig:datasets} with yellow square markers, each obtained by averaging over $M=2000$ realizations of the random circuit, from which it is clear the normalized entanglement $\widetilde{S}_n$ again follows the same behavior of the previous scenarios}.

Thus, the entangling speed can be used as a good estimate of the entanglement generated in a QNN also in real use cases, especially at the start of optimisation, when trainable parameters are usually initialised at random. For example, one could measure the entangling speed of the architecture of interest on a random quantum circuit of just a few qubits, and then estimate the entanglement generated with the same architecture on an arbitrary number of qubits and circuit layers, especially in regimes where simulations are no longer computationally feasible.  

\subsection{Expressibility}\label{sec:Expressibility}
In addition to entanglement, another useful quantity to characterize parametrized quantum circuits is the expressibility, as defined by authors in~\cite{SimPQCs2019}. Such measure quantifies how well the QNN is able to explore the Hilbert space by comparing the distribution of fidelities of states generated by the QNNs with that of randomly Haar-distributed ones (see Appendix~\ref{app:expressibility} for a formal definition and explanation).

\begin{figure}[t]
    \centering
    \includegraphics{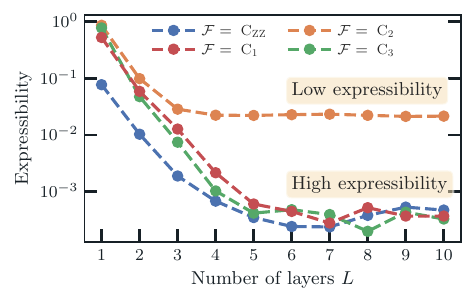}
    \caption{Expressibility of the QNNs analyzed in Fig.~\ref{fig:ent_scaling}, for $n=8$ qubits with linear entanglement. The expressibility measures how well a variational circuit is able to address the unitary space (the lower, the better). All QNNs use the same variational form $\varans = \circuit{2}$, but with different feature maps. As the number of layers is increased, QNNs become more expressible, eventually reaching a plateau.}
    \label{fig:expressibility}
\end{figure}

Thus, in order to have a comprehensive understanding of the factors at play in the behavior of QNNs, in Fig.~\ref{fig:expressibility} we show the expressibility measure for the QNNs analyzed in Fig.~\ref{fig:ent_scaling} with a linear connectivity. As one would expect, the expressibility increases as the number of layers is increased, up until a plateau is reached.

Interestingly, the structure with $\fmap = \varans = \circuit{2}$ turns out to be the least expressible of all the structures considered, even if it is the one producing  entanglement at the fastest rate, in agreement with the results reported in~\cite{SimPQCs2019}, as such QNN is indeed very similar to the parameterized circuit labeled `15' in~\cite{SimPQCs2019}. On the contrary, the QNN with $\fmap = \circuit{zz}$, and $\varans = \circuit{2}$ proposed in~\cite{AbbasPowerQNN2021} is able to reach high expressibility while producing entanglement at a controlled pace. As the presence of high entanglement is correlated with trainability issues~\cite{MarreroBarrenEntanglement2021}, this QNN attains an optimal balance of mild entanglement with high expressibility even at low depth, which could be related to its good performances in quantum machine learning task~\cite{AbbasPowerQNN2021, Havlicek2019QSVM}. However, a similar, yet less favorable balance, is achieved by the other two architectures, so further investigation is needed to discriminate where the optimality comes from. 

In this respect, the authors in~\cite{hubregtsenEvaluationParameterizedQuantum2021} found the expressibility to be correlated with the classification accuracy of QNNs in supervised learning tasks, while weak correlation was found with the entanglement generated inside the circuit, in line with the observations regarding entanglement-induced barren plateaus~\cite{MarreroBarrenEntanglement2021}. As discussed earlier in Sec.~\ref{par:2des_ent_bp}, both expressibility and high entanglement are related to the resemblance of the circuit to a random unitary, but while the former provides a more direct evidence, the latter gives an indirect indication. Indeed, there are cases of circuits having low expressibility but high entanglement, indicating that such circuits selectively explore only some highly-entangled regions of the Hilbert space~\cite{SimPQCs2019}.

\subsection{Distribution of the singular values}\label{sec:mp}
\begin{figure}[t]
    \centering
    \includegraphics{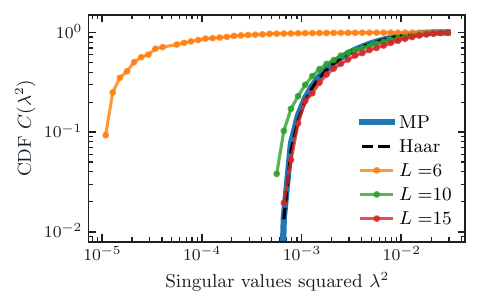}
    \caption{Convergence to the Mar\v{c}enko-Pastur (MP) distribution of the eigenvalues. The cumulative distribution \new{function (CDF)} $C(\lambda)$ of the eigenvalues  $\lambda^2$ corresponding to the central bond of a quantum state of $n=15$ qubits, generated with $\mathcal{F}=\circuit{zz}$, $V=\circuit{2}$, and linear connectivity. We show the behavior for different numbers of layers $L$, and for truly Haar-random states which, as expected, exactly follows the MP curve.}
    \label{fig:mp_convergence}
\end{figure}

The randomness of a quantum state can also be probed using tools from random matrix theory. Specifically, this can be done by studying the distribution of the eigenvalues of the reduced density matrices, which are known to follow the Mar\v{c}enko-Pastur (MP) law when pure random quantum states are considered~\cite{Znidaric_MP_2007, jaschke2022}. More in detail, let $\ket{\psi} \in \mathcal{H}_A \otimes \mathcal{H}_B$ be an Haar-random bipartite quantum state with Schmidt decomposition $\ket{\psi} = \sum_{i=1}^{d} \lambda_i~\ket{\xi_i}_A\otimes\ket{\eta_i}_B$, where $d=\text{min}(d_A, d_B)$ and $d_{A,B}$ is the dimension of the Hilbert space $\mathcal{H}_{A,B}$. The reduced density matrix $\rho_A = \Tr_B\qty[\dyad{\psi}]$ has eigenvalues $\lambda_i^2$ given by the square of the singular values, and for large system size their distribution is described by the MP distribution~\cite{MarcenkoPastur, puchala2016}. 

In Figure~\ref{fig:mp_convergence} we show the cumulative \old{probability distribution} \new{distribution function} of the eigenvalues $C(\lambda^2)$ of the reduced state of the first half of the qubits, obtained with a QNN with $n=15$ qubits, feature map $\mathcal{F} = \circuit{zz}$, variational ansatz $V=\circuit{2}$, and linear connectivity. The distribution of the singular values for the QNN is obtained by running the circuit $10^2$ times with different sets of parameters, and storing the singular values corresponding to the central cut. Then, we construct the cumulative distribution from the histogram of all the singular values obtained from the simulations. As the number of layers $L$ is increased, the distribution of the eigenvalues approaches the theoretical MP distribution, eventually matching it when the number of layers is equal to the number of qubits. This behavior is displayed also by other QNN architectures. For completeness, we also show the distribution of the eigenvalues of a truly Haar-random quantum state, generated by sampling its entries independently from a normal complex distribution and then normalising it~\cite{Znidaric_MP_2007}, which, as expected, follows perfectly the MP curve.

\section{Discussion}\label{sec:Discussion}
Moments of the Haar distribution can be approximated efficiently 
using local random quantum circuits of sufficient depth. Depending on the connectivity dimension $D$ of the qubits, defined as the number of other qubits that are connected to each qubit, order $\mathcal{O}(\text{poly}(t)\cdot n^{1/D})$-depth random circuits are sufficient to create approximate unitary $t$-designs~\cite{BrandaoLocalRandomQuantum2016, Harrow2018approximate, HaferkampRandom, harrowRandomQuantumCircuits2009}, that is circuits that generate a distribution of unitaries which approximately matches moments of the Haar distribution up to order $t$~\cite{Dankert2des2009}. Numerical studies suggest that these results also hold for random parameterized quantum circuits of various forms~\cite{McCleanBarren2018, CerezoBarrenLocalCost2021, Holmes2021connecting, SimPQCs2019}.

We extend these results by showing similar results also for quantum neural networks featuring data re-uploading, both for random instances using random inputs and parameters, and also for real-world dataset when these are used as inputs in the feature map. In particular, for a linear connectivity, as the number of layers approaches to the number of qubits $L \approx n$, QNNs display the same entanglement entropy properties of Haar-distributed random states, a fact which can be taken as a proxy for QNNs approximating unitary designs. Such behaviour was also confirmed by studying the randomness of the circuits with other metrics, namely the expressibility and the convergence to the Mar\v{c}enko-Pastur distribution of the eigenvalues of the reduced states. In both cases, we find strong evidence of the QNN reproducing the same features of random quantum states as the number of layers approaches the system size using a linear connectivity.

Our analysis also underlines the importance of the entangling operations, as careless use of an all-to-all connection can result in unwanted simplifications, making the effective connectivity identical to a nearest-neighbors one. Parameterized two-qubit interactions can solve the problem, even though they may be challenging to implement on real hardware. A good trade-off is achieved with a circular entangling topology, which is immune to simplifications and shows remarkable entangling capabilities. 
Indeed, from the results of Fig.~\ref{fig:ent_scaling}, we see that such connectivity is able to create high multiparite entanglement between qubits already at shallow depth, and with only minor additional hardware resources compared to the linear connectivity. An all-to-all topology instead reaches typical values for entanglement of random states essentially at constant \old{depth $L \in \order{1}$} \new{number of layers $L \in \order{1}$ --- implying in general $\order{n^2}$ entangling operations ---}, independently of the system size, and the architecture used (when non-trivial feature maps and variational ansätze are used).

While limiting the entanglement inside a quantum neural network may be necessary to ensure its trainability~\cite{MarreroBarrenEntanglement2021}, low entanglement makes the circuit prone to be simulated exactly with an MPS, as discussed in Sec.~\ref{sec:MPSSimulation}. Thus, we envision that a sweet spot should be found in order for QNNs to show signs of quantum advantage: not too high to preclude trainability, not to low to escape triviality.

At last, the introduction of the entangling speed $v_s$~\eqref{eq:ent_speed} can be used as a figure of merit for the entanglement production of a given QNN, independent of the size of the system. 
Indeed, the entangling speed can be studied and assigned to an architecture in the simulable regime (low number of qubits $n$), and then used to estimate the number of layers to achieve a well-determined quantity of the entanglement, for any system size. We also stress that $v_S$ characterizes the most interesting interval of layers in a circuit. As discussed earlier, a value of the entanglement too high might be connected to barren plateaus, underlying the importance of exploring the regime where the entanglement has not saturated yet, and the linear regime still holds. 

We now briefly comment on future interesting investigation directions regarding entanglement and QNNs. The focus of this work was to carefully study the entanglement features of common quantum ansätze, specifically when they are initialized with random parameters and no optimization has yet started. A natural followup is to ask whether entanglement plays any role also during the optimization process, which is at core of variational quantum algorithms. While for some specific variational procedures like QAOA~\cite{DupontQAOAEntanglement} or VQE-based ground state solvers~\cite{Woitzik2020EntProdVQE} one has some knowledge of the structure of the target solution, and hence can infer the behaviour of the entanglement created in the circuit, this is not the case for quantum machine learning tasks, as they are usually very task-dependent. Indeed, current proposals for QML advocate for the use of constrained quantum ansätze specifically tailored to the problem under investigation~\cite{SkolikEquivariant, MeyerSymmetries, LaroccaGQML}, and then one expects the depth of the circuit and the entanglement generated inside it to highly depend on the specific task to be solved, and dataset to fit, either classical or quantum~\cite{SharmaNFLEntanglement}. Moreover, while the use of deep QNN ansätze (with arguably more entanglement) could offer some optimization advantages due to overparametrization~\cite{LaroccaOverparameterisation, KianiUnitariesOverparameterisation, AnschuetzLossTrapsQM}, the emergence of barren plateaus suggests using shallow circuits instead~\cite{CerezoBarrenLocalCost2021, WangNoiseinducedBarrenPlateaus2021}. The characterisation of the role played by entanglement in QNNs, and how it may be leveraged to achieve a quantum advantage over classical methods will be objects of future studies.

\section{Conclusion}\label{sec:Conclusion}
In this paper we discussed in detail the entanglement generated by different promising Quantum Neural Networks (QNNs) when these are initialised with random parameters, and showed that they reproduce the same properties of random quantum states under various measures. 

We employed a Matrix Product States (MPS) simulation of the quantum circuits, which guarantees an easy computation of the entanglement in the circuits, and let us study systems of large system size composed of up to $n=50$ qubits.

We showed that while all the architectures tend to a Haar entanglement distribution for a sufficiently high number of layers, the speed of convergence strongly depends on the specific circuit ansatz. This result highlights the universal behavior of the normalized entanglement production~\eqref{eq:normalised_entanglement} for a given architecture, so we introduced a new measure to characterize a QNN in terms of its entanglement production: the entangling speed~\eqref{eq:ent_speed}. 

Finally, we argued that a trade-off between expressibility and entanglement is the key to a better understanding of QNN performances and an auspicious target for the search of quantum advantage. While high entanglement is a necessary condition to avoid classical simulability, a too-large entanglement is detrimental to the training procedure due to its tight connection with barren plateaus, as discussed in Sec.~\ref{sec:Entanglement}. A promising future direction is to extend the entanglement analysis of QNNs not only at initialization but also during the training procedure~\cite{SackBPShadows2022, Joonho2021EntBP, Woitzik2020EntProdVQE}. These tests would help to understand if QNNs really are a suitable platform for proving quantum advantage.

\section*{Code availability}
All simulations with a number of qubits $n\leq 12$ were performed using Qiskit~\cite{Qiskit}, while larger systems were simulated with Quantum Matcha Tea package~\cite{ballarin2021quantum} available at \url{https://baltig.infn.it/quantum_tea/quantum_tea}. The simulations for the training sections were performed using Pennylane~\cite{bergholm2018pennylane}. All the code for reproducing the results presented here is available at the repository: \url{https://github.com/mballarin97/mps_qnn}.

\section*{Acknowledgment}
All authors thank CINECA for providing the necessary resources for running the MPS simulations on the  High-Performance Computing (HPC) infrastructure GALILEO100, the Quantum Computing and Simulation Center (QCSC) at Padova university, and the ICSC National Center. M.B. thanks Eduardo Gonzalez Lazo for carefully reading the manuscript and his valuable feedback and Daniel Jaschke for valuable discussions. Stefano Mangini thanks Stefan Sack for valuable discussions. 


\bibliographystyle{quantum}
\bibliography{bibliography.bib}



\onecolumn\newpage
\appendix

\section{Lower bound on entanglement entropy for unitary 2-designs\label{app:reny}}
The presented derivation is a straightforward application of known results on the entanglement of random states and properties of Rényi-entropies~\cite{LiuEntanglementQuantumRandomness2018, SackBPShadows2022}. Rényi $\alpha$-entropies of a density operator $\rho$ are defined as
\begin{equation}
    S_\alpha(\rho) = \frac{1}{1-\alpha}\log \Tr[\rho^\alpha]\,,
\end{equation}
where $\lim_{\alpha \rightarrow 1}S_{\alpha} (\rho) = S(\rho)$ is the Von Neumann entropy of Eq.~\eqref{eq:ent_entropy}, and it holds that $S_\beta(\rho) \leq S_\alpha(\rho)$ for $\beta \geq \alpha$. Of particular interest is the Rényi 2-entropy $S_2(\rho) = -\log \Tr[\rho^2]$ depending on the purity $\Tr[\rho^2]$ of the system, which is much easier to computer and it can be used to lower bound the Von Neumann entropy via $S(\rho) > S_2(\rho)$. 

Let $\ket{\psi} \in (\mathbb{C}^2)^{\otimes n}$ be the state of a composite system made of subsystems $A$ and $B$ with dimensions $d_A = 2^{n_A}$ and $d_B = 2^{n - n_A}$, respectively. Suppose $\ket{\psi}$ is a random state $\ket{\psi} = U\ket{\psi_0}$, where $U$ is sampled from an ensemble of unitaries that constitutes at least a unitary 2-design. Then, the average value of the purity of the reduced density matrix $\rho_A = \Tr_B[\dyad{\psi}]$ amounts to~\cite{LiuEntanglementQuantumRandomness2018, SackBPShadows2022}
\begin{equation}
    \mathbb{E}_{\text{2-design}} \Tr[\rho_A^2] = \frac{d_A + d_B}{d_Ad_B+1}\, .
\end{equation}
By the convexity of Rényi-entropies with respect to $\Tr[\rho^\alpha]$, and using Jensen's inequality ($\mathbb{E}~f \geq f~\mathbb{E}$), one can lower bound the average Rényi 2-entropy as $\mathbb{E}_{\text{2-design}}[S_2(\rho_A)] \geq -\log\mathbb{E}_{\text{2-design}}[\rho_A^2]$, hence
\begin{equation}
\label{eq:app_s123}
    \mathbb{E}_{\text{2-design}}[S_2(\rho_A)] \geq - \log \frac{d_A + d_B}{d_A d_B + 1} >  \log d_A - \log\frac{d_A + d_B}{d_B} > \log d_A - 1\,.
\end{equation}
Then, since $S(\rho) \geq S_2(\rho)~\forall~\rho$, taking the expectation value on both sides yields a lower bound on the average Von Neumann entropy of $\rho_A$, namely
\begin{equation}
\label{eq:app_s21}
     \log d_A - 1 < \mathbb{E}_{\text{2-design}}[S_2(\rho_A)] \leq \mathbb{E}_{\text{2-design}}[S(\rho_A)] \leq \log d_A.
\end{equation}
which is the bound shown in Eq.~\eqref{eq:2des_vonn} in the main text.

If the state $\ket{\psi}$ is instead a truly Haar-random state, that is $U$ is sampled from the uniform Haar distribution and not just from a 2-design, the entanglement entropy is given by the Page value of Eq.~\eqref{eq:haar_entanglement} in the main text, which is itself lower bounded by~\cite{HaydenAspectsGenericEntanglement2006}
\begin{equation}
\label{eq:app_s22}
    \mathbb{E}_{\text{Haar}}[S(\rho_A)] > \log d_A - \frac{1}{2}\frac{d_A}{d_B} > \log d_A - \frac{1}{2} \quad\quad\quad (d_A < d_B)\, .
\end{equation}

Summarising, for $d_A < d_B$, putting together the bounds~\eqref{eq:app_s21} and~\eqref{eq:app_s22} one has
\begin{gather}
    \log d_A - 1 < \mathbb{E}_{\text{2-design}}[S(\rho_A)] < \log d_A\\
    \log d_A - \frac{1}{2} < \mathbb{E}_{\text{Haar}}[S(\rho_A)] < \log d_A\,,
\end{gather}
Alternatively, in the limit when the subsystem $B$ is much larger than $A$, $d_B \gg d_A$, then by approximating the logarithm $\log(1+x)\approx x$ in~\eqref{eq:app_s123} one also has
\begin{gather}
    \log d_A - \frac{d_A}{d_B} < \mathbb{E}_{\text{2-design}}[S(\rho_A)] < \log d_A\\
    \log d_A - \frac{1}{2}\frac{d_A}{d_B} < \mathbb{E}_{\text{Haar}}[S(\rho_A)] < \log d_A\,,
\end{gather}

Thus, the entanglement entropy of a state sampled from a 2-design is close to that of a truly Haar-random state, with both achieving near-maximal entanglement. Of course, one also expects the Von Neumann entropy of a general $t$-design to be upper bounded by the Page value, $\mathbb{E}_{t\text{-design}}[S(\rho_A)] < \mathbb{E}_{\text{Haar}}[S(\rho_A)]$, with equality obtained in the limit $t \gg 1$.

\section{Computation of the Haar entanglement distribution}
\label{app:haar}
While Eq.~\eqref{eq:haar_entanglement} is the theoretical definition of the Haar entanglement entropy, it is not possible to exactly compute it, due to the exponential number of terms in the sum. However, it is possible to exploit the similarity of the sum with the harmonic series to obtain a good approximation. First, we denote with $H_n$ the truncated harmonic series:
\begin{align}
    H_n = \sum_{k=1}^n\frac{1}{k}.
\end{align}
Then, we rewrite in a more convenient way the sum in Eq.~\eqref{eq:haar_entanglement}:
\begin{align}
    \sum_{j = d_B + 1}^{d_A d_B} \frac{1}{j} = \sum_{j=1}^{d_Ad_B}\frac{1}{j} - \sum_{j=1}^{d_B}\frac{1}{j}=H_{d_Ad_B}-H_{d_B}.
\end{align}
Using well-known results for the truncated Harmonic series~\cite{harmonic_number}:
\begin{align}
    H_n = \log{n} + \gamma + \frac{1}{2n} - \epsilon_n,
\end{align}
where $\gamma\simeq 0.5772$ is the Euler-Mascheroni constant, and $0\leq \epsilon_n\leq 1/8n^2$. Thus, the correction $\epsilon_n$ goes to zero as the number of terms in the sum $n$ increases, allowing for a meaningful approximation of the value. Using this technique, we are able to estimate the Haar entanglement entropy of a $50$-qubits state with an error of the order $~10^{-16}$.

We now proceed to compute the maximum and average of the distribution with a fixed number of qubits $n$. Using Eq.~\ref{eq:haar_entanglement} and recalling $d_{A(B)}=2^{n_{A(B)}}$, $n_B= n-n_A$, $n_A\in [1, n/2]$ we can write:
\begin{align}
    \mathbb{E}[S(\psi_A)] &= H_{d_Ad_B}-H_{d_B} - \frac{d_A-1}{2d_{B}} \\
    &= H_{2^n} - H_{2^{n-n_A}}-\frac{2^{n_A}-1}{2^{n-n_A+1}} \\
    &= \log 2^{n_A}-\frac{2^{n_A}-1}{2^{n-{n_A}+1}} + O\left(\frac{1}{2^{n-n_A}}\right).
\end{align}
We are now interested in the maximum and average of the distribution. It is easy to see that the maximum is achieved for $n_A=n/2$. In this scenario $2^{n_A} \gg 1$:
\begin{align}
    \max_A\big( \mathbb{E}[S(\psi_A)] \big) = \frac{n}{2}\log 2 - \frac{1}{2} + O\left(\frac{1}{2^{n/2}}\right).
\end{align}
Taking into account that for an $n$ qubit system the maximum of the entanglement entropy is $S=\frac{n}{2}\log 2$ we can state that, in the large $n$ limit, a Haar state presents a maximally entangled bond. 

\section{Triviality of the full entangling map} \label{app:entangling_maps}
\begin{figure*}[t]
    \centering
    \includegraphics[width=\textwidth]{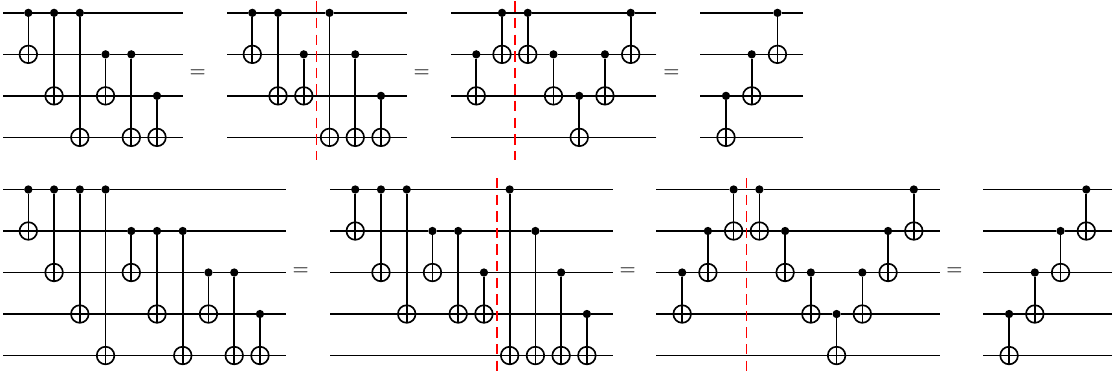}
    \caption{Equivalence of the full entangling map with a nearest-neighbors scheme. Using the circuit identities discussed in the main text, it is straightforward to check that the all-to-all entangling scheme as defined in Alg.~\ref{alg:full} is equivalent to a nearest-neighbors interaction.}
    \label{fig:triviality_full}
\end{figure*}

The full entangling map defined as
\begin{algorithm}[H] 
\caption{Full entangling map}
\label{alg:full}
\begin{algorithmic}[1]
\Require{$q_1, \dots q_n$, qubits} 
\Ensure{Quantum circuit}
\For{$i=1$ to $n$}
    \For{$j=i$ to $n$}
        \State {\textsc{Cnot}$(q_i, q_j)$} 
    \EndFor\EndFor
\end{algorithmic}
\end{algorithm}
can be shown to be equivalent to a nearest neighbors entangling map with the gates in reversed order, see Fig.~\ref{fig:triviality_full}. The proof is straightforward and obtained by direct evaluation, making use of some circuit identities for networks of $\textsc{Cnot}$s~\cite{Garciaescartin2011equivalent}. In particular, (\textit{i}) a \textsc{Cnot} can be distributed into four \textsc{Cnot}s acting on an additional intermediate qubit
\[
\begin{quantikz}[column sep={12.4pt,between origins}, row sep={20pt,between origins}]
& \ctrl{3} & \qw \\
& \qw  		& \qw \\ 
& \qw 		& \qw \\
& \targ{} 	& \qw 
\end{quantikz}
=
\begin{quantikz}[column sep={12.4pt,between origins}, row sep={20pt,between origins}]
& \ctrl{1} 	& \qw 			& \ctrl{1}	& \qw 		& \qw  \\
& \targ{}   	& \ctrl{2}		& \targ{} 	& \ctrl{2} & \qw \\ 
& \qw 			& \qw  			& \qw		& \qw 		& \qw 	\\
& \qw 			& \targ{} 		& \qw		& \targ{} & \qw
\end{quantikz}
\]
(\textit{ii}) \textsc{Cnot}s having different controls and targets commute with each other
\[
\begin{quantikz}[column sep={12.4pt,between origins}, row sep={20pt,between origins}]
& \ctrl{3} 	& \qw 		& \qw\\
& \qw   		& \ctrl{1}	& \qw\\ 
& \qw 			& \targ{}  & \qw\\
& \targ{} 		& \qw 		& \qw
\end{quantikz}
=
\begin{quantikz}[column sep={12.4pt,between origins}, row sep={20pt,between origins}]
& \qw 		& \ctrl{3}	& \qw\\
& \ctrl{1} & \qw		& \qw\\ 
& \targ{} 	& \qw  		& \qw\\
& \qw 		& \targ{} 	& \qw
\end{quantikz}
\]
(\textit{iii}) a cascade of \textsc{Cnot}s can be decomposed as
\[
\begin{quantikz}[column sep={12.4pt,between origins}, row sep={20pt,between origins}]
& \ctrl{3}  & \qw 	 & \qw		& \qw \\
& \qw 		 & \ctrl{2} & \qw		& \qw \\ 
& \qw 		 & \qw		 & \ctrl{1}& \qw \\
& \targ{0}& \targ{} & \targ{}	& \qw 
\end{quantikz}
=
\begin{quantikz}[column sep={12.4pt,between origins}, row sep={20pt,between origins}]
& \ctrl{1} 	& \qw 		& \qw 		& \qw  		& \ctrl{1}& \qw \\
& \targ{}   	& \ctrl{1}	 & \qw 		& \ctrl{1} & \targ{} & \qw \\ 
& \qw 			& \targ{}  & \ctrl{1}	& \targ{} 	& \qw		& \qw \\
& \qw 			& \qw 		& \targ{}	& \qw 		& \qw		& \qw 
\end{quantikz}
\]

The full entangling map can be highly simplified using these three rules, reducing it to a simple sequence of nearest-neighbors interactions. For example, for $n=3$ qubits, using (\textit{i}) to distribute the long-range \textsc{Cnot}, one obtains
\[
\begin{quantikz}[column sep={12.4pt,between origins}, row sep={20pt,between origins}]
& \ctrl{1} & \ctrl{2} & \qw & \qw \\
& \targ{}  & \qw 		& \ctrl{1} & \qw\\ 
& \qw 		& \targ{} &  \targ{} & \qw
\end{quantikz}
=
\begin{quantikz}[column sep={12.4pt,between origins}, row sep={20pt,between origins}]
&  \ctrl{1}\border{2}{2}  	& \ctrl{1} & \qw		& \ctrl{1} & \qw 		 & \qw &  \qw  \\
& \targ{}   	& \targ{}	& \ctrl{1} & \targ{} & \ctrl{1}  & \ctrl{1} &\qw \\ 
& \qw 			& \qw 		& \targ{}	& \qw 	& \targ{} &	\targ{} & \qw
\end{quantikz}
=
\begin{quantikz}[column sep={12.4pt,between origins}, row sep={20pt,between origins}]
& \qw 			& \ctrl{1}	& \qw  \\
& \ctrl{1} 	& \targ{}	& \qw 	\\
& \targ{}   	& \qw 		& \qw 
\end{quantikz}
\]
The simplification process can be iterated for a higher number of qubits by first commuting long range \textsc{Cnot}s at the end of the circuit to create a final cascade, and then making use of the result from the lower dimension case. In Fig.~\ref{fig:triviality_full} the simplification process for $n=4,\,5$ qubits is explicitly shown, and it is directly generalized for all numbers of qubits.

Clearly, these results only hold for networks composed of plain \textsc{Cnot}s, and do not apply for general two-qubit interactions made of controlled unitaries. 

\section{Expressibility of Parameterized Quantum Circuits}\label{app:expressibility}
The expressibility introduced in~\cite{SimPQCs2019} quantifies how well the QNN is able to explore the unitary space by comparing the distribution of fidelities of states generated by the QNN with that of randomly Haar-distributed ones.

Let $U(\bm{\phi})$ be the unitary operation implemented by a parameterized quantum circuit (PQC) with parameters $\bm{\phi}$ (in our case, we would have $\bm{\phi} = (\bm{x},\bm{\theta})$), and be $\ket{\psi_{\bm{\phi}}} = U(\bm{\phi})\ket{\bm{0}}$. Given two realizations of the PQC with parameters $\bm{\phi}_1$ and $\bm{\phi}_2$,  consider the fidelity $F=\abs{\braket{\psi_{\bm{\phi}_1}}{\psi_{\bm{\phi}_2}}}^2$. By repeatedly sampling two sets of parameters and evaluating the corresponding fidelity $F$, one can construct a histogram approximating the probability distribution $\hat{P}_{\text{PQC}}(F)$ of the fidelity for states generated by the PQC. For Haar random quantum states, the probability density function of fidelities is known and amounts to $P_{\text{Haar}}(F) = (N-1)(1-F)^{N-1}$, where $N=2^n$ is the dimension of the Hilbert space~\cite{ZyczkowskiRandomState}. 

The  expressibility is then defined as the Kullback–Leibler divergence $D_{KL}$ between the estimated fidelity distribution and that of a Haar-distributed ensemble, namely
\begin{equation}
    \text{Expressibility} := D_{KL}\qty(\hat{P}_{\text{PQC}}(F) || P_{\text{Haar}}(F))\, .
\end{equation}

\section{Extensive analysis of the entanglement scaling with the increasing depth}
\label{app:ext_ent_scaling}
\begin{figure*}[t]
    \centering
    \includegraphics[width=\textwidth]{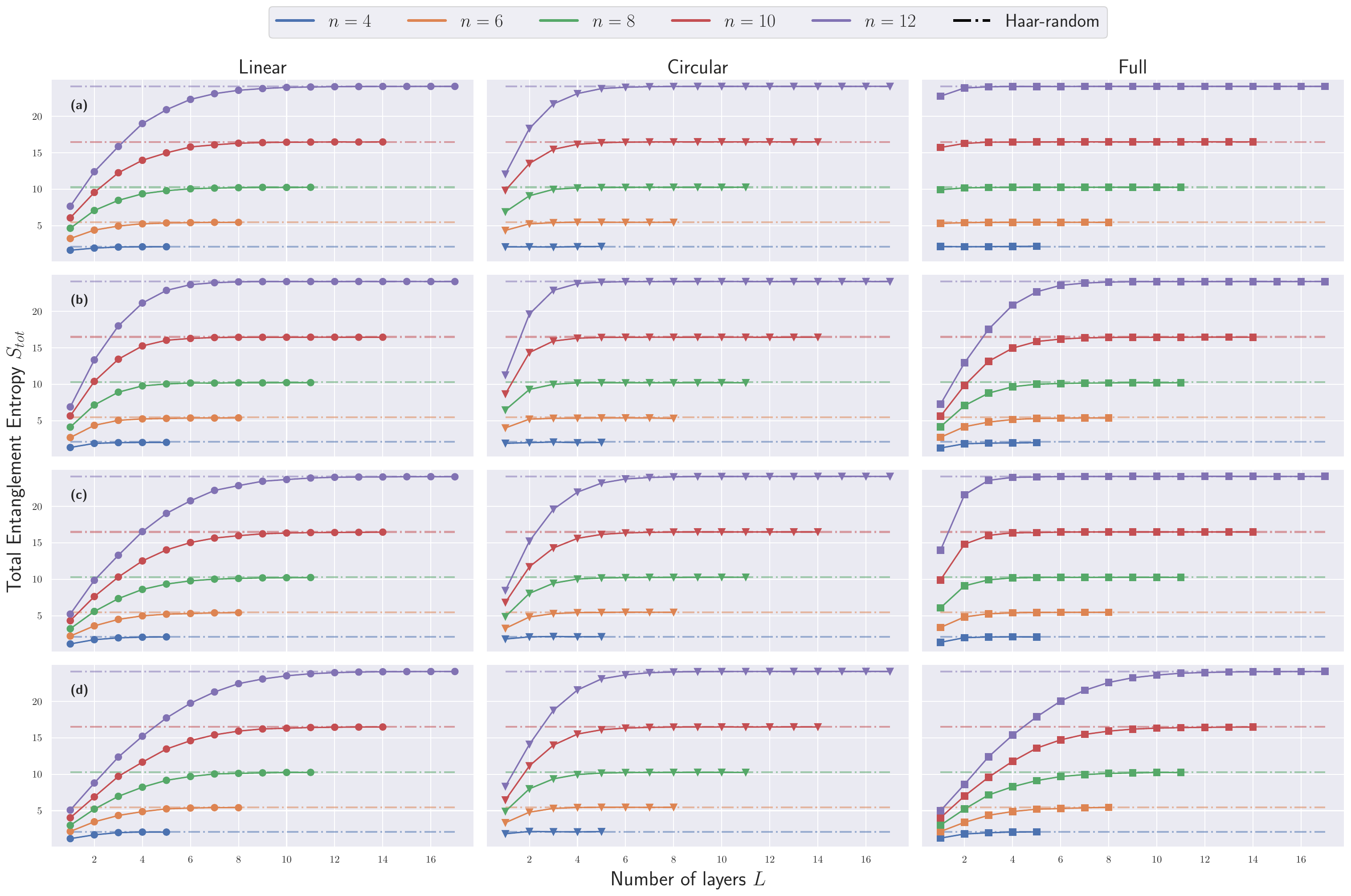}
    \caption{Total entanglement $S_{tot}$~\eqref{eq:total_entanglement} for four different QNN architectures, each evaluated with different entangling topologies (\textit{linear}, \textit{circular} and \textit{full}), shown for increasing number of layers $L$ and for several numbers of qubits $n$. QNNs architecture given by: \textbf{(a)} $\fmap = \circuit{zz}$, $\varans = \circuit{2}$; \textbf{(b)} $\fmap = \circuit{2}$, $\varans = \circuit{2}$; \textbf{(c)} $\fmap = \circuit{3}$, $\varans = \circuit{2}$; \textbf{(d)} $\fmap = \circuit{1}$, $\varans = \circuit{2}$. Note that QNNs leverage the same variational form $\varans$, while the feature map $\fmap$ is changed. See the main text for a discussion of the results.}
    \label{fig:extensive_ent_scaling}
\end{figure*}

In Fig.~\ref{fig:extensive_ent_scaling} we show the behavior of the total entanglement $S_{tot}$ defined in Eq.~\eqref{eq:total_entanglement} for four different QNNs as the depth of the quantum circuit is increased. Note that each QNN is considered with all the three possible entangling topologies (\textit{linear}, \textit{circular} and \textit{full} as defined in Fig.~\ref{fig:main_figure}), and the results are shown for several numbers of qubits $n=4,6,8,10,12$. At last, note that all QNNs leverage the same variational form $\varans = \circuit{2}$, while the feature map is changed, $\fmap = \circuit{zz}, \circuit{2}, \circuit{3}, \circuit{1}$ for panels (a), (b), (c) and (d), respectively. See main text for comments on results.

\section{Convergence of MPS simulations}\label{app:convergence}
Using tensor network, specifically MPS, methods we perform an approximation to simulate large systems, in this work up to $n=50$ qubits. However, the error introduced by the approximations can be monitored, so one always has an estimate of the faithfulness of the tensor network simulation~\cite{jaschke2022}. Let $\ket{\psi_\text{exact}}$ be the true state of the quantum system after the $i$-th two qubit gates in the circuit is applied (one qubit gates do not imply additional approximation errors), and let $\ket{\psi_\text{trunc}}$ denote the truncated quantum state represented by the MPS. The fidelity between these two states evaluated on the $i$-th step of the computation is
\begin{align}
F_i &=|\bra{\psi_\text{exact}}\ket{\psi_\text{trunc}}|^2=\left|\sum_{\alpha=1}^{\chi_\text{exact}}\lambda_\alpha\bra{\xi_\alpha}_1\otimes \bra{\eta_\alpha}_2~\sum_{\beta=1}^{\chi_{s}}\lambda_\beta\ket{\xi_\beta}_1\otimes \ket{\eta_\beta}_2\right|^2 \\
&=\left|\sum_{\alpha=1}^{\chi_{s}} \lambda_\alpha^2\right|^2 = \left|1-\sum_{\alpha=\chi_s+1}^{\chi_\text{exact}}\lambda_\alpha^2\right|^2,
\end{align}
where we represented the states in the Schmidt decomposition with respect to the bond where the $i$-th two-qubit gate was applied, and $\chi_s$ is the bond dimension of the MPS state.
The fidelity $F_t$ of the simulation after application of the $t$-th two-qubit gate is lower bounded by the product of the previous fidelities $F_i$, as~\cite{jaschke2022}
\begin{align}
\label{eq:fid_mps}
    F_t\geq\prod_{i=1}^{t-1} F_i.
\end{align}
where we note that the single step fidelities $F_i$ are readily accessed during the MPS simulation, since one calculates the fidelity before the truncation of the singular values takes place. Equation~\eqref{eq:fid_mps} gives a lower bound to the error introduced by truncation in terms of the fidelity between the true state and the one evolved using an MPS simulation, and one can then control the faithfulness of the simulation at any given time step of the circuit.

\begin{figure}[ht]
    \centering
    \includegraphics{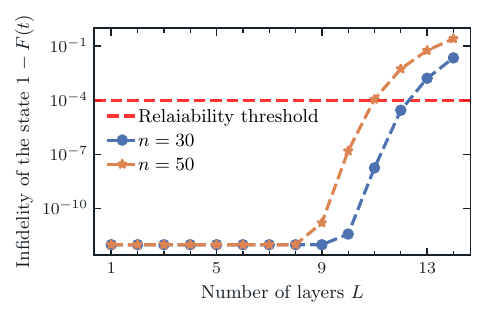}
    \caption{We show the infidelity of the state as a function of the number of layers for $n=30,\, 50$ qubits. The results are reliable up to $L=11$ layers for $n=50$ and up to $L=12$ for $n=30$.}
    \label{fig:mps_convergence}
\end{figure}

In Figure~\ref{fig:mps_convergence} we show the infidelity $1-F$ of the final state from the circuit for $n=30, 50$ with a maximum bond dimension $\chi_s=4096$. The plotted result is the average over $M=10$ realization of the quantum circuit with different sets of parameters. Defining reliable results with the infidelity of at most $1-F=10^{-4}$ we observe that, for $n=50$, we reliably describe circuits up to $11$ layers, while for $n=30$ we can reach $L=12$ layers.

\end{document}